\documentclass[pre,aps,showpacs,twocolumn,floatfix]{revtex4-1}

\usepackage{epsfig}
\usepackage{amsmath}
\usepackage{amsfonts}
\usepackage{amssymb}

\usepackage[normalem]{ulem} 

\usepackage{graphicx}

\begin{document}

\title{Hyperuniformity and Its Generalizations}

\author{Salvatore Torquato}

\affiliation{Department of Chemistry, Department of Physics, Princeton Center for Theoretical Science, Program of Applied and Computational Mathematics, Princeton Institute for the Science and Technology of Materials, Princeton University, Princeton, New Jersey 08544, USA}

\begin{abstract}

Disordered many-particle hyperuniform systems are exotic amorphous states
of matter that lie between a crystal and liquid: they are like perfect crystals in the 
way they suppress large-scale density fluctuations and yet are like liquids or glasses 
in that they are statistically isotropic with no Bragg peaks. 
These exotic states of matter play a vital role in a number of
problems across the physical, mathematical as well as biological sciences and, because
they are endowed with novel physical properties, have
technological importance. Given the fundamental as well as practical importance of disordered hyperuniform systems elucidated thus far,
it is natural to explore the generalizations of the hyperuniformity notion and its consequences. 
In this paper, we substantially broaden the hyperuniformity concept along four different directions.
This includes generalizations to treat fluctuations in the interfacial area (one of the Minkowski
functionals) in  heterogeneous media and surface-area driven evolving microstructures, 
random scalar fields, divergence-free random vector fields, as well as  statistically anisotropic 
many-particle systems and two-phase media. In all cases, the relevant mathematical underpinnings
are formulated and illustrative calculations are provided. Interfacial-area fluctuations play a 
major role in characterizing the microstructure of two-phase systems (e.g., fluid-saturated porous media), physical
properties that intimately depend on the geometry  of the interface, 
and evolving two-phase microstructures that depend on interfacial
energies (e.g., spinodal decomposition). In the instances of
divergence-free random vector fields and  statistically anisotropic structures, we show
that the standard definition of hyperuniformity must be generalized such that it accounts for the dependence 
of the relevant spectral functions on the direction in which the origin in Fourier space
(nonanalyticities at the origin).  Using this analysis, we place some well-known energy spectra from the theory
of isotropic turbulence in the context of this generalization of  hyperuniformity.  
Among other results, we show that there exist  many-particle ground-state configurations
in which directional hyperuniformity imparts exotic  anisotropic physical
properties (e.g., elastic, optical and acoustic characteristics) to these states of matter.
Such tunablity could have technological
relevance for manipulating light and sound waves in ways heretofore not thought possible.
We show that disordered many-particle systems that respond to external fields (e.g., magnetic and electric fields) are a natural
class of materials to look for  directional hyperuniformity. The generalizations of hyperuniformity
introduced here provide theoreticians and experimentalists new research avenues to understand a very broad 
range of phenomena across a variety of fields through the hyperuniformity ``lens."

\end{abstract}
\pacs{05.20.-y,05.40.-a,61.20.Gy,61.50.Ah}

\maketitle

\section{Introduction}


The characterization of density fluctuations in many-body systems is a problem of great fundamental interest in the
physical, mathematical and biological sciences \cite{Wi65,Ka66,Fi67,Wi74,Pe93,Wa96,Te97,No98,Sc99,Chand99,Ga05,Ye03,Mu06,Be07,Ji11d}. 
The anomalous suppression of density fluctuations at very long wavelengths is central 
to the hyperuniformity concept, whose broad importance  for condensed matter physics
and materials science was brought to the fore  only about a decade ago 
in a study that focused on fundamental theoretical aspects, including how it provides
a unified means to classify and categorize crystals, quasicrystals and special
disordered point configurations \cite{To03a}.  Hyperuniform systems are poised at an exotic critical point
in which the direct correlation function, defined via the Ornstein-Zernike relation \cite{Ha86}, is long-ranged \cite{To03a}, in 
diametric contrast to standard thermal and magnetic critical points in which the total correlation function is long-ranged \cite{Wi65,Ka66,Fi67,Wi74}.
Roughly speaking, a hyperuniform many-particle system in $d$-dimensional Euclidean space
$\mathbb{R}^d$ is one in which (normalized)
density fluctuations are completely suppressed at very large length scales,
implying that the structure factor $S({\bf k})$ tends to zero as the wavenumber $k\equiv |\bf k|$ tends to zero,
i.e.,
\begin{equation}
\lim_{|{\bf k}| \rightarrow 0} S({\bf k}) = 0.
\label{hyper}
\end{equation}
Equivalently, it is one in which the number variance $\sigma^2_{_N}(R)$ of particles within a
spherical observation window of radius $R$ grows more slowly than the window volume in the large-$R$ limit, i.e., 
slower than $R^d$. Typical disordered systems, such as liquids and structural glasses, have the standard volume scaling, that is,  $\sigma^2_{_N}(R) \sim R^d$. By contrast, all perfect crystals and quasicrystals are hyperuniform with the surface-area 
scaling $\sigma^2_{_N}(R)\sim R^{d-1}$. Surprisingly,
there are a special  class of disordered particle configurations that have the same asymptotic behavior as crystals.
There are hyperuniform scalings  other than surface-area growth.  When the structure factor
goes to zero in the limit $|{\bf k}| \rightarrow 0$ with the power-law form
\begin{equation}
S({\bf k}) \sim |{\bf k}|^\alpha,
\label{power}
\end{equation}
where $\alpha >0$, the number variance has the following large-$R$ asymptotic scaling \cite{To03a,Za09,Za11b}:
\begin{eqnarray}  
\sigma^2_{_N}(R) \sim \left\{
\begin{array}{lr}
R^{d-1}, \quad \alpha >1\\
R^{d-1} \ln R, \quad \alpha = 1 \qquad (R \rightarrow \infty).\\
R^{d-\alpha}, \quad 0 < \alpha < 1.
\end{array}\right.
\label{sigma-N-asy}
\end{eqnarray}

\begin{figure}
\begin{center}
\includegraphics*[  width=2.in,clip=keepaspectratio]{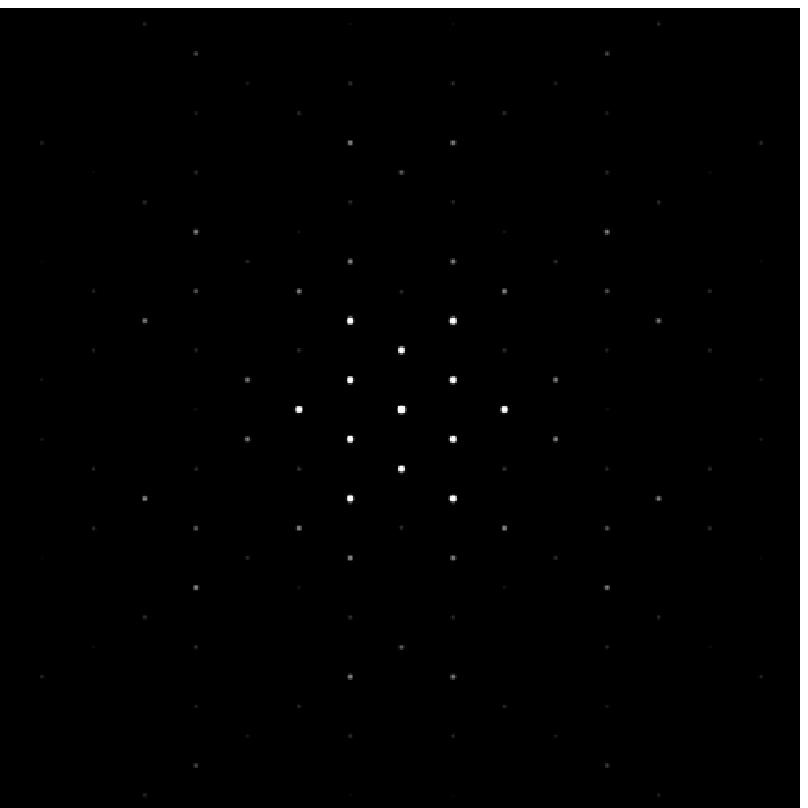}\vspace{0.3in}
 \includegraphics*[  width=2.in,clip=keepaspectratio]{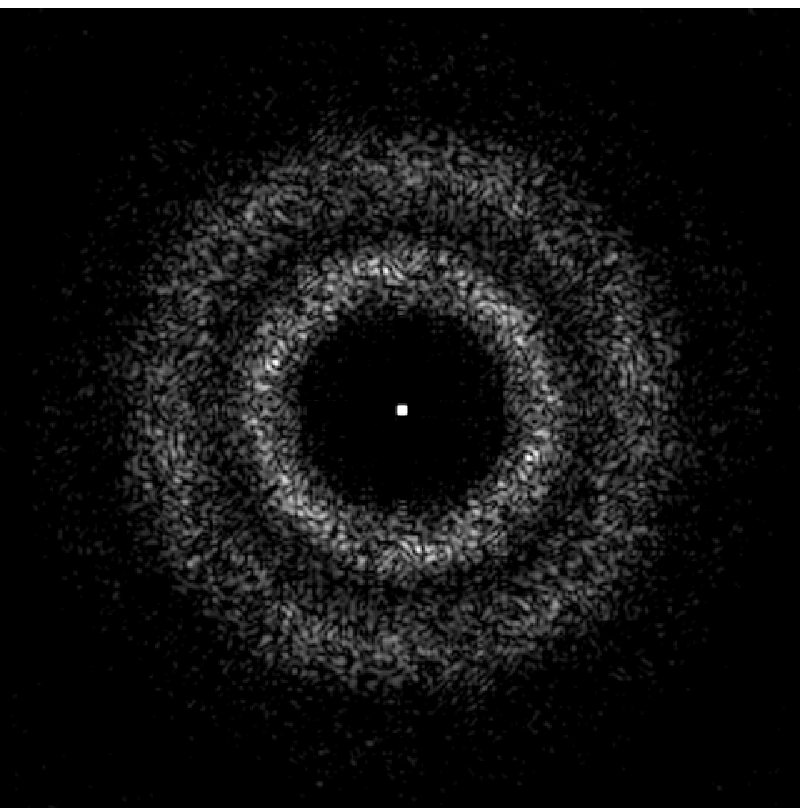}
\caption{Top: Scattering pattern for a crystal. Bottom: Scattering pattern for a  disordered ``stealthy" hyperuniform material \cite{Uc04b,Ba08}.
Notice that apart from forward scattering, there is a circular region around the origin 
in which there is no scattering, a highly exotic situation for an amorphous state of matter.}
\label{pattern}
\end{center}
\end{figure}

Disordered hyperuniform systems are  exotic states
of matter that lie between a crystal and liquid: they are like perfect crystals in the way they suppress large-scale density fluctuations and yet are like liquids or glasses in that they are statistically isotropic with no Bragg peaks and  hence 
have no long-range order. In this sense, they can have a {\it hidden order  on large length scales} (see Fig. 2 of Ref. \cite{To15} for a vivid example) and, because of their hybrid nature,
are endowed with novel physical properties, as described below. Figure \ref{pattern} shows a typical scattering pattern
for a crystal and another for a ``stealthy" disordered hyperuniform one in which there is a circular region around the
origin in which there is no scattering and diffuse scattering outside this ``exclusion" zone \cite{Uc04b,Ba08}, a highly unusual for an amorphous material.

We knew only  a few examples of {\it disordered} hyperuniform systems (also known as ``superhomogeneous" patterns)
about a decade ago \cite{Leb83,To03a,Ga02}.
 We now know that these exotic states of matter can exist as {\it equilibrium} and {\it nonequilibrium} phases, 
of both  the classical and the quantum-mechanical varieties. Examples include
``stealthy" disordered  ground states \cite{Uc04b,Ba08,To15,Zh15a,Zh15b},
maximally random jammed particle packings \cite{Do05d,Za11a,Ji11c,Ch14a},  jammed athermal granular media~\cite{Be11}, jammed thermal colloidal packings~\cite{Ku11,Hu12,Dr15}, dynamical processes in ultracold atoms~\cite{Le14}, 
disordered networks with large photonic band gaps \cite{Fl09b}, driven nonequilibrium systems 
\cite{He15,Ja15,We15,Tj15,Sc15,Di15}, avian photoreceptor patterns \cite{Ji14}, geometry of neuronal tracts \cite{Bur15},
immune system receptors \cite{Ma15}, certain quantum ground states (both fermionic and bosonic) \cite{To08c,Fe56}, high-density transparent materials
\cite{Le16},  and wave dynamics in disordered potentials \cite{Yu15}. Hyperuniformity  has pointed to new correlation functions
from which one can extract relevant growing length scales as a function of temperature
as a liquid is supercooled below its glass transition temperature \cite{Ma13a,Co16},
a problem of great interest in glass physics \cite{Lu07,Be07,Sc07,Ka09,Chand10,Hock12}.
Remarkably, the one-dimensional point patterns derived from the nontrivial zeros
of the Riemann zeta function \cite{Mon73} and the eigenvalues of random Hermitian matrices \cite{Dy62a}
are disordered and hyperuniform.

A variety of groups
have recently fabricated disordered hyperuniform materials at the
micro- and nano-scales for various photonic applications \cite{Man13a,Man13b,Ha13},
surface-enhanced Raman spectroscopy \cite{De15}, 
realization of a terahertz quantum cascade laser \cite{Deg15}, 
 and self-assembly of diblock  copolymers \cite{Zi15b}.  Moreover, it was
shown that the electronic band gap of amorphous silicon
widens as it tends toward a hyperuniform state \cite{He13}. Recent
X-ray scattering  measurements indicate that amorphous-silicon samples
can be made to be nearly hyperuniform   \cite{Xie13}.

The hyperuniformity concept was generalized to the case of heterogeneous materials \cite{Za09}, i.e.,
materials consisting of two or more phases \cite{Note1}.
Heterogeneous materials abound in Nature and synthetic
situations. Examples include composite and porous media, biological media (e.g., plant and animal tissue), 
foams, polymer blends, suspensions,
granular media, cellular solids, and colloids \cite{To02a}. In the case of two-phase media
(defined more precisely in Sec. \ref{back}), one relevant fluctuating quantity is the local phase volume fraction
within a window. The simplest characterization of such fluctuations is the local volume-fraction
variance  $\sigma_{_V}^2(R)$ associated with a $d$-dimensional spherical window of radius $R$ \cite{Lu90b,Qu97b,To02a,Note2}.
It was demonstrated  that  the hyperuniformity condition in the context of
volume-fraction fluctuations in a two-phase heterogeneous system   is
one in which the variance  $\sigma_{_V}^2(R)$ for large $R$ goes to zero more 
rapidly than the inverse of the window volume  \cite{Za09}, i.e., faster than $R^{-d}$, which is equivalent to the following condition
on the relevant spectral density ${\tilde \chi}_{_V}({\bf k})$ (defined in Sec. \ref{back}):
\begin{eqnarray}
\lim_{|\mathbf{k}|\rightarrow 0}\tilde{\chi}_{_V}(\mathbf{k}) = 0.
\label{hyper-2}
\end{eqnarray}
This generalization of the hyperuniformity concept
has been fruitfully applied to characterize a variety of disordered two-phase systems \cite{Za11a,Za11c, Za11d,Dr15,Ch15},
and the rational design of digitized hyperuniform two-phase media with tunable
disorder \cite{Di16}.
As in the case of hyperuniform point configurations \cite{To03a,Za09,Za11b}, 
it is easily shown that three different scaling regimes 
arise  in the case of hyperuniform two-phase systems when the spectral density
goes to zero with the power-law form ${\tilde \chi}_{_V}({\bf k})\sim |{\bf k}|^\alpha$: 
\begin{eqnarray}  
\sigma^2_{_V}(R) \sim \left\{
\begin{array}{lr}
R^{-(d+1)}, \quad \alpha >1\\
R^{-(d+1)} \ln R, \quad \alpha = 1 \qquad (R \rightarrow \infty).\\
R^{-(d+\alpha)}, \quad 0 < \alpha < 1
\end{array}\right.
\label{sigma-V-asy}
\end{eqnarray}

\begin{figure}[bthp]
\centerline{\includegraphics[  width=2.4in, keepaspectratio,clip=]{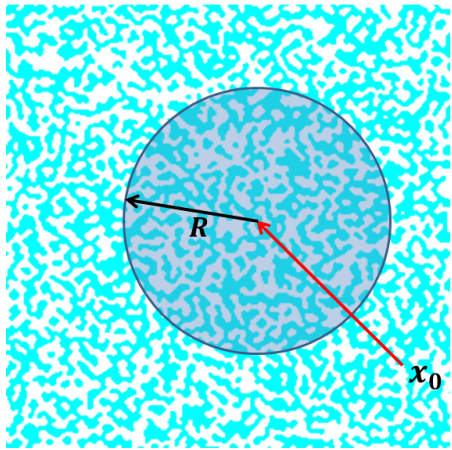}}
\caption{(Color online) A schematic indicating a circular observation window  of radius $R$ that is centered at
position $\bf x_0$ in a disordered two-phase medium; one phase is depicted
as a blue (darker) region and the other phase as a white region. The phase volume fractions or interfacial
area within the window will fluctuate as the window position ${\bf x}_0$ is varied. }
\label{patterns}
\end{figure}

Given the fundamental as well as practical importance of disordered hyperuniform systems elucidated thus far,
it is natural to explore further generalizations of the hyperuniformity notion and its consequences. 
In this paper, we  extend the hyperuniformity concept in a variety of different directions.
Before doing so, 
we make some  remarks about hyperuniformity in two-phase systems in which
one phase is a  sphere packing (Sec. \ref{packing-1}).  
We then introduce the notion of hyperuniformity
as it concerns local fluctuations in the interfacial area in disordered two-phase media and apply
the mathematical formulation to sphere packings (Sec. \ref{area}). We demonstrate
that surface-area fluctuations are considerably more sensitive
microstructural measures than volume-fraction fluctuations, and hence
provide a more powerful approach to understand
hyperuniformity in two-phase systems.
Subsequently, we extend the  hyperuniformity concept to random scalar fields
(Sec. \ref{scalar}). Such phenomena are ubiquitous and include, but  are not limited
to, concentration and temperature
fields in heterogeneous  media and turbulent  flows,
laser speckle patterns, and temperature fluctuations associated
with the cosmic microwave background.
Among other results, we show how a random scalar field can inherit the hyperuniformity
property from an underlying hyperuniform point process.
We also note that the analysis for continuous random fields 
is trivially extended to discrete cases derived from 
experimental images or computer-simulation studies.
We then generalize the hyperuniformity formalism to treat random vector fields
and find that this extension requires one to broaden the definition
of hyperuniformity to account for the dependence of the relevant spectral
tensor function on the direction in which the origin 
is approached (Sec. \ref{vector}). Mathematically, this means that the directional-dependent
spectral tensor associated with a hyperuniform vector field is nonanalytic at the origin.
This is to be contrasted with previous definitions of hyperuniformity, which assumed that the
way in which the origin in Fourier space (scatterling pattern)
is approached is independent of direction.  Generalizing the definition of hyperuniformity
to account for directionality  provides completely new and potentially exciting avenues for theoretical and experimental work, 
including the possibility to
design random vector fields  with targeted hyperuniform spectra. 
Among other results, we reinterpret and analyze well-known turbulent energy spectra in the context
of this generalization of hyperuniformity.  Subsequently, the notion of directional hyperuniformity is proposed 
in the context of many-particle systems and heterogeneous media
that are statistically anisotropic (Sec. \ref{aniso}). Here we show that
directionality in Fourier space can again play a pivotal role. In particular,
directional hyperuniformity imparts exotic  anisotropic physical
properties (e.g., elastic, optical and acoustic characteristics) to these states of matter.
Finally, we offer concluding remarks and a discussion (Sec. \ref{con}).

\section{Definitions and Background}
\label{back}

\subsection{Point Configurations}
\label{points}

Consider $N$ points with configuration
${\bf r}^N \equiv {\bf r}_1,{\bf r}_2,\ldots,{\bf r}_N$ in a large region $\cal V$ of volume $V$ 
in  $d$-dimensional Euclidean space $\mathbb{R}^d$.
Any single point configuration is specified by its {\it microscopic density} $n({\bf r})$ at
position $\bf r$, which is a random variable defined by
\begin{equation}
n({\bf r})=\sum_{j=1}^N \delta({\bf r}-{\bf r}_j),
\label{local-den}
\end{equation}
where $\delta({\bf r})$ is a $d$-dimensional Dirac delta function.
The point process is statistically characterized by the {\it specific}
probability density function $P_N({\bf r}^N)$, where $P_N({\bf r}^N)d{\bf
r}^N$ gives the probability of finding point 1 in volume element
$d{\bf r}_1$ about ${\bf r}_1$, point 2 in volume element
$d{\bf r}_2$ about ${\bf r}_2$, $\ldots$, and point $N$  in volume element
$d{\bf r}_N$ about ${\bf r}_N$. Thus, $P_N({\bf r}^N)$ normalizes
to unity and $d{\bf r}^N \equiv d{\bf r}_1 d{\bf
r}_2\cdots d{\bf r}_N$ represents the $(Nd)$-dimensional volume element. 
The ensemble average of any function $f({\bf r}^N)$ that depends
on the point configuration ${\bf r}^N$ is given by
\begin{equation}
\langle f({\bf r}^N) \rangle = \int_{\cal V} \int_{\cal V} \cdots \int_{\cal V} f({\bf r}^N)
P_N({\bf r}^N)d{\bf r}^N.
\label{ensemble}
\end{equation}
The reduced {\it generic}
density  function $\rho_n({\bf r}^n)$ ($n <N$), defined as
\begin{eqnarray}
\rho_n({\bf r}^n) = \frac{N!}{(N-n)!}  \int_V \cdots \int_V P_N({\bf r}^N) d{\bf r}^{N-n},
\label{rhon}
\end{eqnarray}
where $d{\bf r}^{N-n}\equiv d{\bf r}_{n+1} d{\bf r}_{n+2} \cdots  d{\bf r}_N $.
The quantity $\rho_n({\bf r}^n)d{\bf r}^n$ is proportional to
the probability of finding {\it any} $n$ particles ($n \le N$)
with configuration $\bf r^n$ in volume element $d{\bf r}^n$.

For statistically homogeneous media, $\rho_{n}({\bf r}^n)$
is translationally invariant and hence depends only on the relative
displacements, say with respect to ${\bf r}_1$:
\begin{equation}
\rho_{n}({\bf r}^n)=\rho_{n}({\bf r}_{12},{\bf r}_{13},\ldots,{\bf r}_{1n}),
\end{equation}
where ${\bf r}_{ij}={\bf r}_j-{\bf r}_i$. The one-particle
function $\rho_1$ is just equal to the constant {\it number density}
\index{number density} of particles $\rho$, i.e.,
\begin{equation}
\rho_1({\bf r}_1) = \rho \equiv \lim_{ N,V\rightarrow\infty} \frac{N}{V} .
\label{thermolimit}
\end{equation}
This limit  is referred to as  the
{\it thermodynamic limit}.  It is convenient to define the so-called
{\it $n$-particle correlation function}
\begin{equation}
g_n({\bf r}^n) = \frac{\rho_n({\bf r}^n)}{\rho^n}.
\label{nbody}
\end{equation}
In the absence of long-range order and when the
particles are mutually far from one another (i.e.,  ${r}_{ij}=|{\bf r}_{ij}|
\rightarrow\infty$,
$1\leq i < j \leq N$), $\rho_n({\bf r}^n) \rightarrow \rho^n$ and 
$g_n({\bf r}^n) \rightarrow 1$. 

The important two-particle quantity
\begin{equation}
g_2({\bf r}_{12}) = \frac{\rho_2({\bf r}_{12})}{\rho^2}
\label{g2-rho2}
\end{equation}
is usually referred to as the {\it pair correlation function}.
The {\it total correlation function} $h({\bf r}_{12})$ is defined as
\begin{equation}
h({\bf r}_{12})=g_2({\bf r}_{12})-1,
\label{total}
\end{equation}
which is trivially related to the autocovariance function associated
with the random variable (\ref{local-den}), i.e,
\begin{equation}
 \frac{1}{\rho}\Bigg\langle    \Big(n({\bf x})- \rho \Big) \, \Big(n({\bf x}+{\bf r}) -  \rho\Big) \,\Bigg\rangle = \delta({\bf r})
+ \rho h({\bf r})
\label{H}
\end{equation}
where we have invoked statistical homogeneity.

Spectral representations of direct-space pair statistics of various types
are central to the hyperuniformity concept. 
We use the following definition of the
Fourier transform of some function $f({\bf r})$, which can represent a {\it tensor
of arbitrary rank} and depends on the 
vector $\bf r$ in  $\mathbb{R}^d$:
\begin{eqnarray}
        \tilde{f}(\mathbf{k}) = \int_{\mathbb{R}^d} f(\mathbf{r}) \exp\left[-i(\mathbf{k}\cdot  \mathbf{r})\right] d\mathbf{r},
\end{eqnarray}
where $\mathbf{k}$ is a wave vector. 
When it is well-defined, the corresponding inverse Fourier transform is given by
\begin{eqnarray}
f(\mathbf{r}) = \left(\frac{1}{2\pi}\right)^d \int_{\mathbb{R}^d}       \tilde{f}(\mathbf{k}) \exp\left[i(\mathbf{k}\cdot  \mathbf{r})\right] d\mathbf{k}.
\end{eqnarray}
If  $f$ is a radial function, i.e., depends 
on the modulus $r=|\mathbf{r}|$ of the vector $\bf r$, 
its Fourier transform is given by
\begin{eqnarray}
{\tilde f}(k) =\left(2\pi\right)^{\frac{d}{2}}\int_{0}^{\infty}r^{d-1}f(r)
\frac{J_{\left(d/2\right)-1}\!\left(kr\right)}{\left(kr\right)^{\left(d/2\right
)-1}} \,d r,
\label{fourier}
\end{eqnarray}
where  $k=|{\bf k} |$ is wavenumber or modulus of the wave vector $\bf k$
and $J_{\nu}(x)$ is the Bessel function of order $\nu$.
The inverse transform of $\tilde{f}(k)$ is given by
\begin{eqnarray}
f(r) =\frac{1}{\left(2\pi\right)^{\frac{d}{2}}}\int_{0}^{\infty}k^{d-1}\tilde{f}(k)
\frac{J_{\left(d/2\right)-1}\!\left(kr\right)}{\left(kr\right)^{\left(d/2\right
)-1}} d k.
\label{inverse}
\end{eqnarray}
We recall the first several terms in  the series expansion of $J_{\nu}(x)$
about $x=0$:
\begin{eqnarray}
\hspace{-0.15in}J_{\nu}(x) &=&  \frac{(x/2)^{\nu}}{\Gamma(\nu +1)}- \frac{(x/2)^{\nu+2}}{\Gamma(\nu +2)}+ \frac{(x/2)^{\nu+4}}{2\Gamma(\nu +3)} -{\cal O}(x^{\nu +6}),\nonumber\\
\end{eqnarray}
which we will apply later in the article.

The nonnegative structure factor $S(\bf k)$ is the Fourier transform
of the autocovariance function (\ref{H}) and is trivially related to ${\tilde h}({\bf k})$,
which is the Fourier transform of the total correlation function $h(\bf r)$:
\begin{equation}
S({\bf k})=1+\rho {\tilde h}({\bf k}).
\label{factor}
\end{equation}
The structure factor is proportional to the scattering intensity.  It is useful to recall
the relationship between the local number variance $\sigma^2_N(R)$ associated
with a spherical window of radius $R$ for
a point configuration \cite{To03a}:
\begin{eqnarray}
\sigma_N^2(R)&=& \rho v_1(R)\Big[1+\rho  \int_{\mathbb{R}^d} h({\bf r}) 
\alpha(r;R) d{\bf r}\Big] \nonumber \\
&=&
 \rho v_1(R)\Big[\frac{1}{(2\pi)^d} \int_{\mathbb{R}^d} S({\bf k})
{\tilde \alpha}(k;R) d{\bf k}\Big],
\label{local}
\end{eqnarray}
where 
\begin{equation}
v_1(R) =\frac{\pi^{d/2} R^d}{\Gamma(1+d/2)}
\label{v1}
\end{equation}
is the volume of a $d$-dimensional sphere of radius $R$, and
$\alpha(r;R)$ is the {\it scaled intersection volume}, the ratio of the intersection volume of two spherical windows
of radius $R$ whose centers are separated by a distance $r$ to the volume of
a spherical window, known analytically in any space dimension \cite{To02a,To06b}. Its Fourier transform 
is  given by
\begin{equation}
{\tilde \alpha}(k;R)= 2^d \pi^{d/2} \Gamma(1+d/2)\frac{[J_{d/2}(kR)]^2}{k^d},
\label{alpha-k}
\end{equation}
which clearly is a nonnegative function. Here $J_{\nu}(x)$ is the Bessel function of order $\nu$.

The hyperuniformity condition (\ref{hyper})
defined through the structure factor and relation (\ref{local}) implies that
the number variance $\sigma^2_N(R)$ grows more slowly than $R^d$ for large $R$.
Observe that  hyperuniformity requirement (\ref{hyper}) dictates that the 
volume integral of  $\rho h({\bf r})$ over all space is exactly
equal to $-1$, i.e.,
\begin{equation}
\rho \int_{\mathbb{R}^d} h({\bf r}) d{\bf r}=-1,
\label{sum-1}
\end{equation}
which can be thought of as a sum rule. 
{\it Stealthy} configurations are those in which
the structure factor $S({\bf k})$ is exactly zero for a subset of wave vectors, meaning that they completely suppress
single scattering of incident radiation for those wave vectors. 
{\it Stealthy hyperuniform} patterns \cite{Uc04b,Ba08,To15} are a subclass of hyperuniform
systems in which $S({\bf k})$ is zero for a range
of wave vectors around the origin, i.e.,
\begin{equation}
S({\bf k})= 0 \qquad \mbox{for}\; 0 \le |{\bf k}| \le K,
\label{stealth}
\end{equation}
where $K$ is some positive number.
 An example of a disordered  stealthy  and hyperuniform scattering pattern
is shown in the bottom panel of Fig. \ref{pattern}.

\subsection{Two-Phase Media}
\vspace{-0.1in}

A two-phase  random medium is a domain of space $\mathcal{V} \subseteq \mathbb{R}^d$ of volume $V$
that is partitioned into two disjoint regions: a 
phase 1 region $\mathcal{V}_1$ 
and A phase 2 region $\mathcal{V}_2$ 
such that $\mathcal{V}_1 \cup \mathcal{V}_1 =\mathcal{V}$ \cite{To02a}. 
Denote by $\partial {\mathcal V}$ the interface between  $\mathcal{V}_1$ and $\mathcal{V}_2$.\vspace{-0.25in}

\subsubsection{Phase Statistics}\vspace{-0.1in}

The phase indicator function ${\cal I}^{(i)}({\bf x})$ for a given realization is defined as
\begin{equation}
{\cal I}^{(i)}({\bf x}) = \left\{
{\begin{array}{*{20}c}
{1, \quad\quad {\bf x} \in {\cal V}_i,}\\
{0, \quad\quad {\bf x} \notin {\cal V}_i},
\end{array} }\right.
\label{phase-char}
\end{equation}
\noindent 
The one-point correlation function $S_1^{(i)}({\bf x})= \langle {\cal I}^{(i)}({\bf x}) \rangle$
(where angular brackets indicate an ensemble average) is independent of position $\bf x$,  
for statistically homogeneous media, namely, the constant phase volume fraction, i.e.,
\begin{equation}
\phi_i = \langle {\cal I}^{(i)}({\bf x}) \rangle.
\end{equation}
The two-point correlation function is defined as $S^{(i)}_2({\bf x}_1,{\bf x}_2) = \left\langle{{\cal I}^{(i)}({\bf x}_1){\cal I}^{(i)}({\bf x}_2)}\right\rangle$,
 This function is the probability
of finding two points ${\bf x}_1$ and ${\bf x}_2$  in phase $i$, and for homogeneous  media,  
 depends only on the relative displacement vector ${\bf r} \equiv {\bf x}_2-{\bf x}_1$ 
and hence $S_2^{(i)}({\bf x}_1,{\bf x}_2)=S_2^{(i)}({\bf r})$. 
The autocovariance function $\chi_{_V}({\bf r})$ associated with the random variable ${\cal I}^{(i)}({\bf x})$ 
is given by
\begin{equation}
\label{eq108}
\chi_{_V}({\bf r}) \equiv  S^{(1)}_2({\bf r}) - \phi^2_ 1 =  S^{(2)}_2({\bf r}) - \phi^2_2.
\end{equation}
The nonnegative spectral density ${\tilde \chi}_{_V}({\bf k})$, which can be obtained from  scattering experiments \cite{De49,De57},
is  the Fourier transform of $\chi_{_V}({\bf r})$.
Higher-order versions of these correlation functions
\cite{To82b,To83a,To02a} (not considered here)  arise in rigorous bounds
and exact expressions for effective transport \cite{To85f,Be85a,Be88b,Se89,Gi95a,To02a,Mi02,Ph03,To04a}, 
elastic \cite{Be88b,Gi95a,To97b,To02a,Mi02} and electromagnetic \cite{Re08a}
properties of two-phase media.

It is known that the volume-fraction variance $\sigma_{V}^2(R)$
within a $d$-dimensional spherical window of radius $R$ can be expressed in terms of the autocovariance function $\chi_{_V}({\bf r})$ \cite{Lu90b} or of the spectral density ${\tilde \chi}_{_V}(\mathbf{k})$:
\begin{eqnarray}
\sigma_{_V}^2(R) &=& \frac{1}{v_1(R)} \int_{\mathbb{R}^d} \chi_{_V}(\mathbf{r}) \alpha(r; R) d\mathbf{r} \nonumber \\
&=& \frac{1}{v_1(R)(2\pi)^d} \int_{\mathbb{R}^d} {\tilde \chi}_{_V}(\mathbf{k}) {\tilde \alpha}(k; R) d\mathbf{k},
\label{phi-var-2}
\end{eqnarray}
where, as in relation (\ref{local}), $\alpha(r;R)$ is the scaled intersection volume of two
spherical windows, and ${\tilde \alpha}(k; R)$ is its Fourier transform.
The  hyperuniformity requirement (\ref{hyper-2}) dictates that the autocovariance
function $\chi_{_V}({\bf r})$ exhibits both positive and negative correlations such that
its volume integral over all space is exactly zero, i.e.,
\begin{equation}
\int_{\mathbb{R}^d} \chi_{_V}({\bf r}) d{\bf r}=0,
\label{sum-2}
\end{equation}
which can be regarded to be a sum rule.

We note in passing that realizability conditions for the existence of hyperuniform
autocovariances and spectral densities of general
two-phase media have recently been explored \cite{To16b}. These conditions
restrict the possible class of functional forms that can be hyperuniform. \vspace{-0.25in}

\subsubsection{Interfacial Statistics}\vspace{-0.1in}

The interface between the phases of a realization of a two-phase medium
is generally known probabilistically and is characterized by 
the interface indicator function ${\cal M}({\bf x})$ \cite{To02a} defined as
\begin{equation}
{\cal M}({\bf x})= |\nabla {\cal I}^{(1)}({\bf x})|=|\nabla {\cal
I}^{(2)}({\bf x})|
\label{surf-char}
\end{equation}
and therefore is a {\it generalized} function that is nonzero when
$\bf x$ is on the interface.  The specific surface $s$
(interface area per unit volume) is a one-point correlation
given by the expectation of ${\cal M}({\bf x})$: 
\begin{equation}
s= \langle {\cal M}({\bf x}) \rangle,
\label{s(x)}
\end{equation}
where, because of the assumption of statistical homogeneity,  $s$ is independent of the position $\bf x$.

One can define a variety of higher-order surface correlation functions \cite{To02a}, but for
our purposes in this paper, we will restrict ourselves to the following two-point correlation
function:
\begin{equation}
F_{ss}({\bf r}) = \left\langle{{\cal M}({\bf x}){\cal M}({\bf x} +{\bf r})}\right\rangle,
\label{surface}
\end{equation}
which is called the {\it surface-surface} correlation function. Note the definition (\ref{surface}) invokes 
the statistical homogeneity of the process. The surface-surface correlation function
arises in rigorous bounds on the effective rate constants for diffusion-controlled
reactions \cite{Do76,Ru88} and fluid permeability \cite{Do76,Ru89a} of fluid-saturated porous media. The autocovariance associated
with the random variable ${\cal M}$ for homogeneous media
is given by
\begin{equation}
 \chi_{_S}(\mathbf{r}) = F_{ss}({\bf r}) - s^2,
\label{auto-S}
\end{equation}
which, unlike the dimensionless autocovariance $\chi_{_V}({\bf r})$, has dimensions of  inverse of length squared,
independent of the dimension $d$.
The nonnegative spectral density ${\tilde \chi}_{_S}({\bf k})$ is  the Fourier transform of $\chi_{_S}({\bf r})$,
when it exists. \vspace{-0.1in}


\section{Some Remarks About Two-Point Statistics and Hyperuniform Sphere Packings}
\label{packing-1}\vspace{-0.1in}

Here we collect various known results scattered throughout
the literature concerning the autocovariance function
$\chi_{_V}({\bf r})$ and spectral density ${\tilde \chi}_{_V}({\bf k})$ for two-phase
media in $\mathbb{R}^d$ in which one phase is a sphere packing in order to compare them 
to corresponding results for the surface-surface correlation function and the generalization of hyperuniformity
to surface-area fluctuations introduced in the subsequent section.

A particle packing is a configuration of nonoverlapping (i.e., hard)
particles in $\mathbb{R}^d$.
For statistically homogeneous packings of congruent spheres  of radius $a$ in $\mathbb{R}^d$ at number density $\rho$,
the two-point probability function $S_2({\bf r})$ of the particle (sphere) phase is known exactly in terms of the pair correlation function \cite{To85b,To02a}; specifically,
\begin{eqnarray}
{\chi}_{_V}({\bf r}) &=& \rho\, m_v(r;a) \otimes m_v(r;a) +\rho^2 m_v(r;a) \otimes m_v(r;a) \otimes h({\bf r}) \nonumber \\
&=& \rho \,v_2^{int}(r;a) +\rho^2 v_2^{int}(r;a) \otimes h({\bf r}),
\label{S2-spheres}
\end{eqnarray}
where \vspace{-0.35in}

\begin{equation}
m_v(r;a) =\Theta(a-r)=\Bigg\{{1, \quad r \le a,\atop{0, \quad r > a,}}
\label{indicator}
\end{equation}
is a spherical particle indicator function  \cite{Note3}.
$\Theta(x)$ is the Heaviside step-function,
and $v_2^{int}(r;a)=v_1(a)\alpha(r;a)$ is the intersection volume of two spheres
of radius $a$ whose centers are separated by a distance $r$, where $v_1(a)$ and $\alpha(r;a)$
are defined as in (\ref{phi-var-2}), and $\otimes$ denotes the convolution of two
functions $F({\bf r})$ and $G({\bf r})$:\vspace{-0.27in}

\begin{equation}
F({\bf r}) \otimes G({\bf r}) =\int_{\mathbb{R}^d} F({\bf x}) G({\bf r}-{\bf x}) d{\bf x}.
\end{equation}
Fourier transformation of (\ref{S2-spheres}) gives the spectral
density in terms of the structure factor \cite{To85b,To02a,Za09}:\vspace{-0.25in}

\begin{eqnarray}
{\tilde \chi}_{_V}({\bf k})&=& \rho \,{\tilde m}^2(k;a)+ \rho^2 {\tilde m}^2(k;a) {\tilde h}({\bf k}) \nonumber \\
&=& \rho\, {\tilde m}^2(k;a) S({\bf k})  \nonumber \\
&=& \phi {\tilde \alpha}(k;a) S({\bf k}) 
\label{chi_V-S}
\end{eqnarray}
\vspace{-0.35in}

\noindent{where}\vspace{-0.3in}

\begin{equation}
\hspace{-0.1in}{\tilde \alpha}(k;a)= \frac{1}{v_1(a)} {\tilde m}^2(k;a)= \frac{1}{v_1(a)} \left(\frac{2\pi a}{k}\right)^{d} J_{d/2}^2(ka),
\end{equation}
and
\vspace{-0.35in}

\begin{equation}
\phi =\rho v_1(a),
\end{equation}
is the {\it packing fraction}.

Using relation (\ref{chi_V-S}), it follows that the hyperuniformity of a sphere packing 
can only arise  if the underlying point configuration (sphere
centers) is itself hyperuniform, i.e., ${\tilde \chi}_{_V}({\bf k})$ inherits the hyperuniformity property (\ref{hyper-2})
only through the structure factor, not ${\tilde \alpha}(k;a)$; see Ref. \cite{To16b} for more details.
The stealthiness property, i.e., no scattering at some finite subset of wave vectors (Sec. \ref{points}),
is a bit more subtle.  Relation (\ref{chi_V-S}) dictates
that ${\tilde \chi}_{_V}({\bf k})$ is zero at those wave vectors where $S({\bf k})$ is zero as well as
at the zeros
of the function ${\tilde \alpha}(k;a)$, which is determined by the zeros of the Bessel function
$J_{d/2}(ka)$. The function  ${\tilde \chi}_{_V}({\bf k})$  will be zero at all
of the zeros of ${\tilde \alpha}(k;a)$ for any disordered packing free of any Dirac delta functions (Bragg peaks),
hyperuniform or not.

These results for the pair statistics in direct and Fourier spaces 
have been generalized to the case of impenetrable spheres
with a size distribution at overall number density $\rho$ \cite{Lu91,To02a}.
The  Supplemental Material describes these equations as they concern hyperuniformity \cite{Note4}.

\section{Interfacial Area Fluctuations and Hyperuniformity}
\label{area}\vspace{-0.1in}

Here we  introduce the idea of hyperuniformity associated with local fluctuations in the interfacial area of two-phase media in $\mathbb{R}^d$
and derive the relevant formulas. This generalization provides new tools to analyze 
a variety of phenomena that occur in physical and biological systems
in which interfaces play a dominant role. For example, the geometry of the interface in a fluid-saturated
porous medium  is crucial in determining the fluid permeability \cite{Do76,Ru89a}
and  trapping rate \cite{Do76,Ru88} associated with diffusion and reaction in such systems.   
Another striking class of examples include surface-energy driven coarsening phenomena,
such as those that occur in spinodal decomposition and morphogenesis \cite{Ca58,Sw77}.\vspace{-0.2in}

\subsection{Local Specific-Surface Fluctuations}\vspace{-0.1in}

While the global specific surface  defined by (\ref{s(x)}) is a fixed constant,  the specific surface on a local scale
 determined by an observation window clearly fluctuates, as in the case of the local phase volume fraction. 
Here we derive an explicit expression for the variance associated with the local specific surface and the corresponding hyperuniformity condition.
For simplicity, we consider a $d$-dimensional spherical window of radius $R$ centered
at position ${\bf x}_0$ (see Fig. \ref{patterns}) for statistically homogeneous
two-phase media. The associated \emph{local dimensionless specific surface} $\tau_{_S}(\mathbf{x}_0;R)$ within 
a window of radius $R$ centered at position ${\bf x}_0$ is specified
explicitly by 
\begin{eqnarray}\label{one}
\tau_{_S}(\mathbf{x}_0; R) = \frac{1}{s v_1(R)}\int {\cal M}(\mathbf{x}) w(\mathbf{x}-\mathbf{x}_0; R) d\mathbf{x},
\label{tau}
\end{eqnarray}
where $v_1(R)$ is given by (\ref{v1}), ${\cal M}(\mathbf{x})$ is the interface indicator
function defined by (\ref{surf-char}),  $s$ is the specific surface given by (\ref{s(x)}), and $w$ is the corresponding window indicator function
defined by 
\begin{equation}
w({\bf r};R)=\Bigg\{{1, \quad |{\bf r}| \le R,\atop{0, \quad  |{\bf r}| > R.}}
\label{window}
\end{equation}
Notice that in the limit $R \rightarrow \infty$, the dimensionless random variable $\tau_{_S}(\mathbf{x}_0; R)$
tends to unity.
The variance $\sigma_{S}^2(R)$ associated with fluctuations in dimensionless specific surface is defined by
\begin{eqnarray}\label{three}
\sigma^2_{_S}(R) &\equiv& \langle\tau_{_S}^2(\mathbf{x}_0; R) \rangle -  \langle\tau_{_S}(\mathbf{x}_0; R) \rangle^2 \nonumber\\
&=& \langle\tau_{_S}^2(\mathbf{x}_0; R) \rangle - 1,
\label{def}
\end{eqnarray}
where we have used the fact that the ensemble average  $\langle\tau_{_S}(\mathbf{x}_0; R) \rangle=1$,
which is independent of the window position ${\bf x}_0$ because the system is statistically homogeneous.

Substitution of (\ref{tau}) into (\ref{def}) yields
\begin{eqnarray}
\hspace{-0.3in}\sigma^2_{_S}(R) &=&\frac{1}{s^2 v_1^2(R)} \Big[\int F_{ss}({\bf r}) w(\mathbf{x}_1-\mathbf{x}_0; R)  \nonumber \\
 && \qquad \times w(\mathbf{x}_2-\mathbf{x}_0; R) d\mathbf{x}_1d\mathbf{x}_2\Big] -1,
\end{eqnarray}
where ${\bf r}={\bf x}_2 -{\bf x}_1$.
Using the definition of the scaled intersection volume of two windows of radius $R$,
\begin{equation}
\alpha(r;R)= \frac{1}{v_1(R)} \int_{\mathbb{R}^d}  w(\mathbf{x}_1-\mathbf{x}_0; R) w(\mathbf{x}_2-\mathbf{x}_0; R) d\mathbf{x}_0,
\end{equation}
and the identity \cite{To03a}
\begin{equation}
\frac{1}{v_1(R)} \int_{\mathbb{R}^d} \alpha(r;R) d{\bf r}=1
\end{equation}
leads to the desired relation for the local specific-surface variance:
\begin{eqnarray}
\sigma_{_S}^2(R) = \frac{1}{s^2 v_1(R)} \int_{\mathbb{R}^d} \chi_{_S}(\mathbf{r}) \alpha(r; R) d\mathbf{r},
\label{s-var-1}
\end{eqnarray}
where $\chi_{_S}(\mathbf{r})$ is the autocovariance function associated with the interface 
indicator function [cf. (\ref{auto-S})], $r=|\bf r|$, and we have invoked statistical homogeneity.
The alternative  Fourier representation of the surface-area variance
that is dual to the direct-space representation (\ref{s-var-1}) is trivially obtained 
by applying Parseval's theorem to (\ref{s-var-1}),  provided that the 
spectral density ${\tilde \chi}_{_S}({\bf k})$ exists:
\begin{eqnarray}
\sigma_{_S}^2(R) = \frac{1}{s^2 v_1(R)(2\pi)^d} \int_{\mathbb{R}^d} {\tilde \chi}_{_S}(\mathbf{k}) {\tilde \alpha}(k; R) d\mathbf{k}.
\label{s-var-2}
\end{eqnarray}

A two-phase system is hyperuniform with respect to surface-area fluctuations if the spectral density ${\tilde \chi}_{_S}({\bf k})$
obeys the condition
\begin{eqnarray}
\lim_{|\mathbf{k}|\rightarrow 0}\tilde{\chi}_{_S}(\mathbf{k}) = 0,
\label{hyper-3}
\end{eqnarray}
which implies the sum rule
\begin{equation}
\int_{\mathbb{R}^d} \chi_{_S}({\bf r}) d{\bf r}=0.
\label{sum-3}
\end{equation}
This hyperuniformity property  is equivalent to requiring that the surface-area variance  $\sigma_{S}^2(R)$ for large $R$ goes to zero more rapidly  than $R^{-d}$, which is the same condition as that for the volume-fraction variance  discussed in the Introduction.
Using precisely the same analysis as for point configurations \cite{To03a,Za09,Za11b}, 
it is simple to show that three different hyperuniform scaling regimes 
arise from (\ref{s-var-2}) when the surface-area spectral density
goes to zero with the power-law form ${\tilde \chi}_{_S}({\bf k}) \sim |{\bf k}|^\alpha$: 
\begin{eqnarray}  
\sigma^2_{_S}(R) \sim \left\{
\begin{array}{lr}
R^{-(d+1)}, \quad \alpha >1\\
R^{-(d+1)} \ln R, \quad \alpha = 1 \qquad (R \rightarrow \infty).\\
R^{-(d+\alpha)}, \quad 0 < \alpha < 1
\end{array}\right.
\label{sigma-S-asy}
\end{eqnarray}
Note that these scaling forms are exactly the same as those for volume-fraction fluctuations
[cf.  (\ref{sigma-V-asy})].

\subsection{Sphere Packings}\vspace{-0.1in}
\label{packing-2}

Here we make some remarks about hyperuniformity  associated with specific-surface 
fluctuations in the case of sphere packings. To do so, we first must collect some known
results for their interfacial two-point statistics.
In the special instance of packings of congruent spheres  of radius $a$ in $\mathbb{R}^d$ at number density $\rho$,
the  autocovariance function $\chi_{_S}({\bf r})$ is known exactly in terms of the pair correlation function \cite{To86i,To02a}:
\begin{equation}
{\chi}_{_S}({\bf r}) = \rho\, m_s(r;a) \otimes m_s(r;a)  +\rho^2 m_s(r;a) \otimes m_s(r;a)  \otimes h({\bf r}),
\label{F-spheres}
\end{equation}
where 
\begin{equation}
m_s(r;a)= \frac{\partial m_v(r;a)}{\partial a}= \delta(r-a) ,
\label{delta}
\end{equation}
is a interface indicator function for a sphere,
$\delta(r)$ is a radial Dirac delta function, and $m(r;a)$ is  defined
by (\ref{indicator}). Note that the first term on the right side
of relation (\ref{F-spheres}), which has support in the interval $[0,2a]$, generally possesses an integrable singularity at the origin \cite{To86f}.
Fourier transformation of (\ref{F-spheres}) gives the corresponding spectral
density in terms of the structure factor \cite{To86f,To02a}:
\begin{equation}
{\tilde \chi}_{_S}({\bf k})=\rho\, {\tilde m}_s^2(k;a) S({\bf k}),
\label{chi_S-S}
\end{equation}
where ${\tilde m}_s(k;a)$ is the Fourier transform of the radial Dirac delta function (\ref{delta}) given by
\begin{equation}
{\tilde m}_s(k;a)=\frac{\partial {\tilde m}_v(k;a)}{\partial a}=\left(\frac{2\pi a}{k}\right)^{d/2} k \, J_{d/2-1}(ka).
\end{equation}
The global specific surface $s$, defined generally by (\ref{s(x)}), is given by
\begin{equation}
s= \rho {\tilde m}_s(k=0;a) = \rho s_1(a) = \frac{d\phi}{a},
\end{equation}
where
\begin{equation}
s_1(a) \equiv \frac{\partial v_1(a)}{\partial a}= \frac{d \pi^{d/2} a^{d-1}}{\Gamma(1+d/2)},
\end{equation}
is the surface area of a $d$-dimensional sphere of radius $a$. Thus, since ${\tilde m}_s(k;a)$ is a positive well-behaved function in the vicinity of  $k=0$, it immediately follows from expression (\ref{chi_S-S}) that if the underlying
point process is hyperuniform and/or stealthy,  then
the spectral density ${\tilde \chi}_{_S}({\bf k})$ inherits the same hyperuniformity property (\ref{hyper-3}).
More generally, relation (\ref{chi_S-S}) requires that the spectral density ${\tilde \chi}_{_S}({\bf k})$  is zero at those wave vectors where $S({\bf k})$ is zero (or stealthy) and
at the zeros of the function ${\tilde m}_s(k;a)$.

To compare volume-fraction and surface-area fluctuations statistics to one another, we consider an example where these quantities  can be calculated exactly for a sphere-packing
model as density increases up to a hyperuniform state. Specifically, we consider
$d$-dimensional sphere packings corresponding to  a  certain  $g_2$-invariant process
introduced by Torquato and Stillinger \cite{To03a}. A $g_2$-invariant process
is one in which a chosen nonnegative form for
the pair correlation function $g_2$ remains
invariant over a nonvanishing density \cite{To02b}. The upper
limiting ``terminal'' density is the point above which
the nonnegativity condition on the structure factor
[cf. (\ref{factor})] would be violated. Thus, whenever the structure
factor attains its minimum value of zero at ${\bf k}=0$ at the terminal
or critical density, the system, if realizable, is hyperuniform.
In Ref. \cite{To03a}, a variety of hyperuniform $g_2$-invariant processes
in which the number variance $\sigma^2_{_N}(R)$ grows like the window surface
area (i.e., $R^{d-1}$) were exactly  studied in arbitrary space dimensions.

For our purposes, we use the ``step-function" $g_2$-invariant process, namely, a
$g_2(r)$ that is defined by the unit step function $\Theta(r-D)$, where
$D=2a$ is the sphere diameter.  It is noteworthy that large particle configurations in one, two and three dimensions
that achieve the step-function $g_2(r)$ for densities
up to the terminal density $\rho_c$ have been numerically constructed  \cite{Cr03,Uc06a}.
Interestingly, the  ``ghost" random-sequential-addition  packing  is an exactly solvable model
with an identical terminal density $\rho_c=[2^d v_1(D/2)]^{-1}$ and  a pair correlation function 
that is very nearly equal to a step function  and indeed exactly approaches the step function
in the large-$d$ limit \cite{To06a}. The  structure factor for the step-function $g_2$-invariant process
 in the density range $0 \le \rho \le \rho_c$ is exactly given by
\begin{equation}
S({\bf k})=1-\Gamma(1+d/2) 
\left(\frac{2}{kD}\right)^{d/2}
\left(\frac{\rho}{\rho_c}\right) J_{d/2}(kD),
\label{invariant}
\end{equation}
where  $\rho_c=[2^dv_1(D/2)]^{-1}$ is the terminal density
at which the packing is hyperuniform \cite{To03a}  with a small-$k$ asymptotic scaling
given by
\begin{equation}
S({\bf k}) = \frac{1}{2(d+2)} (kD)^2 + {\cal O}\left((kD)\right)^4.
\end{equation}
For $\rho < \rho_c$, the packing
is not hyperuniform.  
  Substitution of (\ref{invariant}) into relations (\ref{chi_V-S}) and (\ref{chi_S-S}) yields for this model in $d$ dimensions
the associated spectral densities for the phase volumes and interface, respectively,
\begin{eqnarray}
\hspace{-0.1in}{\tilde \chi}_{_V}({\bf k})&=&\rho \left(\frac{\pi D}{k}\right)^{d} J_{d/2}^2(kD/2)\nonumber \\
&&\times \Bigg[ 1-\Gamma(1+d/2) 
\left(\frac{2}{kD}\right)^{d/2}
\left(\frac{\rho}{\rho_c}\right) J_{d/2}(kD)\Bigg]\nonumber \\
\label{CHI}
\end{eqnarray}
and 
\begin{eqnarray}
\hspace{-0.1in}{\tilde \chi}_{_S}({\bf k})&=&\rho \left(\frac{\pi D}{k}\right)^{d} k^2 J_{d/2-1}^2(kD/2)\nonumber \\
&&\times\Bigg[ 1-\Gamma(1+d/2) 
\left(\frac{2}{kD}\right)^{d/2}
\left(\frac{\rho}{\rho_c}\right) J_{d/2}(kD)\Bigg].\nonumber \\
\label{CHI-S}
\end{eqnarray}
(Note that formula (\ref{CHI}) was reported and studied elsewhere \cite{To16b}.)
At the terminal density $\rho_c$, these spectral functions also go to zero quadratically in $k$
in the limit $k \rightarrow 0$ such that
\begin{equation}
{\tilde \chi}_{_V}({\bf k}) = \frac{1}{2(d+2)4^d v_1(1)} (kD)^2 + {\cal O}\left((kD)\right)^4.
\end{equation}
and
\begin{equation}
{\tilde \chi}_{_S}({\bf k}) = \frac{d^2}{2(d+2)4^{d-1} v_1(1)} (kD)^2 + {\cal O}\left((kD)\right)^4,
\end{equation}
but the latter has a coefficient that grows quadratically faster in the dimension relative 
to that in the former.

\begin{figure}
\begin{center}
\includegraphics[  width=3.in, keepaspectratio,clip=]{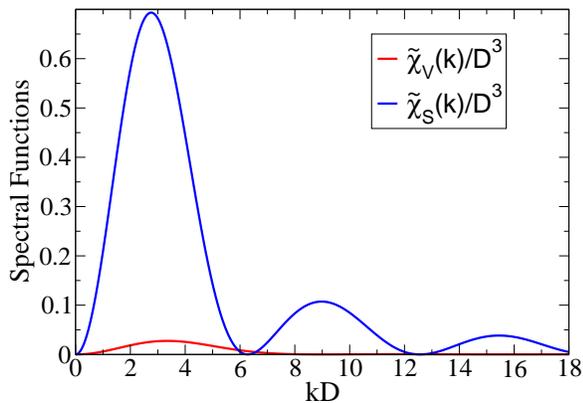}
\caption{(Color online)  Comparison of the two hyperuniform spectral functions ${\tilde \chi}_{_V}(k)$ (lower curve) and ${\tilde \chi}_{_S}(k)$
versus wavenumber $k$ for a sphere packing corresponding to the step-function $g_2$-invariant process in three dimensions at the 
hyperuniform terminal density $\rho_c=3/(4\pi)$ \cite{To03a}.
Here $D$ is the diameter of a hard sphere.}
\label{specs}
\end{center}
\end{figure}

\begin{figure}
\begin{center}
\includegraphics[  width=3.in, keepaspectratio,clip=]{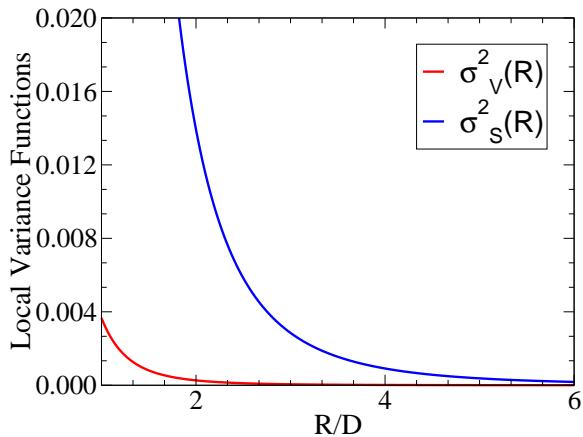}
\caption{(Color online)  Comparison of the volume-fraction variance $\sigma^2_{_V}(R)$ (lower curve) and 
surface-area  variance  $\sigma^2_{_S}(R)$
versus window sphere radius $R$ for a sphere packing corresponding to the step-function $g_2$-invariant process in three dimensions at the 
hyperuniform terminal density $\rho=3/(4\pi)$ \cite{To03a}. Here $D$ is the diameter of a hard sphere.}
\label{vars}
\end{center}
\end{figure}

Figure \ref{specs} shows the two spectral functions, ${\tilde \chi}_{_V}({\bf k})$ and ${\tilde \chi}_{_S}({\bf k})$,
for the step-function $g_2$-invariant packing process in three dimensions at the terminal density $\rho_c=3/(4\pi)$,
as obtained from (\ref{chi_V-S}), (\ref{chi_S-S}) and (\ref{invariant}) with  $a=D/2$.
Figure \ref{vars} depicts the associated local variances for the same system, as obtained
from these spectral functions, and relations (\ref{phi-var-2}) and (\ref{s-var-2}).
Notice that the surface-area spectral function exhibits stronger and longer-ranged correlations 
compared to the volume-fraction spectral function, indicating that the former
is a more sensitive microstructural descriptor.
Figure \ref{vars} depicts the corresponding local variances for the same system.
Similarly, while the corresponding local variances decay like $R^{-4}$ for large $R$,
the surface-area variance does so at a slower rate relative to the volume-fraction
counterpart.

The aforementioned  results for the surface-area pair statistics 
were generalized to the case of sphere packings
with a continuous or discrete size distribution \cite{Lu91,To02a}.
These results are collected in Appendix A  in order to
describe the conditions under which they are  ``multihyperuniform."

\section{Random Scalar Fields and Hyperuniformity}\vspace{-0.1in}
\label{scalar}

Here we generalize the hyperuniformity concept to random scalar fields in $\mathbb{R}^d$.
Such fields can arise in a variety of physical contexts, including concentration and temperature
fields in heterogeneous and porous media \cite{To02a,Sa03} as well as in turbulent  flows \cite{Ba59,Mo75},
laser speckle patterns \cite{Pi88,Wi08,Dog15,Dib16}, and temperature fluctuations associated
with the cosmic microwave background \cite{Pe93,Kom03}. Other example include spatial patterns that arise
in biological and chemical systems that have been theoretically described by, for example,
Cahn-Hilliard \cite{Ca58} and Swift-Hohenberg equations \cite{Sw77}.
In what follows, we derive the relevant equations to quantify
hyperuniform scalar fields, present
illustrative calculations, and remark on two-phase media that
result from level cuts.

\subsection{Local Field Fluctuations}\vspace{-0.1in}

Consider a statistically homogeneous random scalar field $F(\bf x)$ in $\mathbb{R}^d$ that is real-valued with an autocovariance function
\begin{equation}
\psi({\bf r})= \Bigg\langle    \Big(F({\bf x}_1)- \langle F({\bf x})_1\rangle\Big) \, \Big(F({\bf x}_2) -  \langle F({\bf x}_2)\rangle\Big) \,\Bigg\rangle,
\label{spec-field}
\end{equation}
where we have invoked the statistical homogeneity of the field, since ${\bf r}={\bf x}_2 -{\bf x}_1$, which is a $d$-dimensional vector. We assume that 
the associated spectral density ${\tilde \psi}({\bf k})$ (Fourier
transform of the autocovariance) exists. The hyperuniformity condition
is simply that the nonnegative spectral density obeys the 
small-wavenumber condition:
\begin{equation}
\lim_{|{\bf k}| \rightarrow {\bf 0}} {\tilde \psi}({\bf k})=0,
\label{hyp-field}
\end{equation}
which implies the sum rule
\begin{equation}
\int_{\mathbb{R}^d} \psi({\bf r}) d{\bf r}=0.
\end{equation}
The local variance associated with fluctuations in the field, denoted by $\sigma_{_F}^2(R)$, 
is related to the autocovariance function or spectral function  in the usual way:
\begin{eqnarray}
\sigma^2_{_F}(R)&=& \frac{1}{v_1(R)} \int_{\mathbb{R}^d} \psi(\mathbf{r}) \alpha(r; R) d\mathbf{r},\nonumber \\
&=&\frac{1}{ v_1(R)(2\pi)^d} \int_{\mathbb{R}^d} {\tilde \psi}({\bf k})
{\tilde \alpha}(k;R) d{\bf k}.
\label{local-scalar}
\end{eqnarray}

While the main focus of this section is continuous random scalar fields,
it should be noted that when simulating random fields on the computer or when
extracting them from experimentally obtained images,
one must inevitably treat discrete or digitized  renditions of the fields.
The ``pixels" or "voxels" (smallest components of the digitized systems
in 2D and 3D dimensions, respectively) take on
gray-scale intensities that span the intensity range 
associated with the continuous field. Thus, the discrete versions
of relations (\ref{spec-field}) and (\ref{local-scalar}) are to be applied
in such instances; see, for example, Ref. \cite{Bl93}.

\subsection{Random Fields Derived from Point Configurations}\vspace{-0.1in}

Now we prove that a class of fields derived from underlying hyperuniform point configurations
are themselves hyperuniform.
Consider a general ensemble of point configurations of $N$ points
in a large region of volume $V$ in $\mathbb{R}^d$.
Let $K({\bf x};{\bf C})$ represent a nonnegative dimensionless scalar  kernel function that is
radial in $\bf x$ and sufficiently localized so that its Fourier transform exists. Here $\bf C$ represents a set
of parameters that characterizes the shape of the radial function.
Following Blumenfeld and Torquato \cite{Bl93}, the random scalar field $F(\bf x)$ 
is defined as a convolution of the microscopic density and the kernel, i.e.,
\begin{eqnarray}
F({\bf x}) &= &\int_{\mathbb{R}^d} n({\bf x}^\prime) K({\bf x}-{\bf x}^\prime) d{\bf x}^\prime \nonumber\\
&=& \sum_{i=1}^N K({\bf x}-{\bf r}_i)
\label{field}
\end{eqnarray}
where we have dropped indicating the explicit dependence of the kernel on the
parameter set $\bf C$.
It is seen that the effect of the kernel is to smooth out the point ``intensities."
Ensemble averaging (\ref{field}) and using the definition (\ref{ensemble}), yields
the expectation of the field:
\begin{eqnarray}
\langle F({\bf x}) \rangle &=& \left< \sum_{i=1}^N K({\bf x}-{\bf
r}_i) \right> \nonumber \\
&=& \int_V \int_V \cdots \int_V \sum_{i=1}^N K({\bf x}-{\bf
r}_i) P_N({\bf r}^N) d{\bf r}^N \nonumber \\
&=&  \int_V \rho_1({\bf r}_1) K({\bf x}-{\bf r}_1)  d{\bf r}_1 \nonumber \\
&=&  \rho \int_{\mathbb{R}^d}  K({\bf x}) d{\bf x}, 
\label{N(R)}
\end{eqnarray}
where in the last line we have invoked the statistical homogeneity of the field and hence have taken the thermodynamic
limit.
Similarly, the autocorrelation function associated with the field is given by
\begin{eqnarray}
\hspace{-0.3in}\langle F({\bf x}) F({\bf x}+{\bf r}) \rangle &=& \left < \sum_{i=1}^N K({\bf x}-{\bf r}_i)  K({\bf x}+{\bf r}-{\bf r}_i)\right>  \nonumber \\
&+& \left< \sum_{i\neq j}^N K({\bf x}-{\bf r}_i)  K({\bf x}+{\bf r}-{\bf r}_j) \right>  \nonumber \\
&=&\rho K({\bf r}) \otimes K({\bf r}) \nonumber \\  &+& \rho^2   K({\bf r})  \otimes K({\bf r})  \otimes  h({\bf r}) 
+ \langle F \rangle^2,
\end{eqnarray}
where  $h({\bf r})$ is the total correlation function
for the point configuration defined by (\ref{total}).
Thus, the autocovariance function $\psi({\bf r})$, defined generally by (\ref{spec-field}), is given by
\begin{eqnarray}
\psi({\bf r}) = \rho K({\bf r}) \otimes K({\bf r}) + \rho^2   K({\bf r}) \otimes K({\bf r}) \otimes  h({\bf r}).
\label{chi-F}
\end{eqnarray}
Fourier transforming (\ref{chi-F}) yields the corresponding nonnegative spectral density:
\begin{equation}
{\tilde \psi}({\bf k})=\rho {\tilde K}^2({\bf k}) S({\bf k}),
\label{chi-field}
\end{equation}
where ${\tilde K}({\bf k})$ is the Fourier transform of the kernel $K({\bf x})$
and $S({\bf k})$ is the ensemble-averaged structure factor [cf. (\ref{factor})].
We see from (\ref{chi-field}) that if the underlying point process is hyperuniform,
i.e., $S({\bf k})$ tends to zero in the limit $|{\bf k}|\rightarrow 0$, and ${\tilde K}({\bf k})$
is well-behaved at $\bf k=0$, the spectral
density obeys the hyperuniformity condition (\ref{hyp-field}).

\begin{figure}[bthp]
\begin{center}
\includegraphics[  width=3.in, keepaspectratio,clip=]{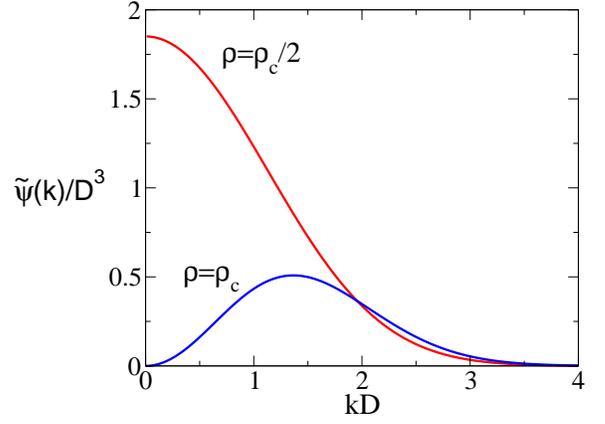}
\caption{ (Color online) 
The spectral function ${\tilde \psi}({\bf k})$ versus wavenumber $k$ for the three-dimensional Gaussian field derived from   
the step-function $g_2$-invariant packing for a nonhyperuniform case ($\rho=\rho_c/2$) and 
the unique hyperuniform instance ($\rho=\rho_c$). Here $\rho_c=3/(4\pi)$  and $a=D$, where $D$ is a hard-sphere diameter.}
\label{gauss}
\end{center}
\end{figure}

As a simple example, consider the Gaussian kernel function:
\begin{equation}
K({\bf r}) =\exp(-(r/a)^2)
\end{equation}
where $a$ is a characteristic length scale that is proportional to the standard
deviation of the Gaussian. The corresponding Fourier transform  is given by
\begin{equation}
{\tilde K}({\bf k}) =\pi^{d/2} a^d\exp[-(ka)^2/4].
\end{equation}
Consider the hyperuniform structure factor (\ref{invariant}) 
for the step-function $g_2$-invariant packing.  Substitution of (\ref{invariant}) into relation (\ref{chi_V-S}) yields
the associated spectral density for this model in $d$ dimensions:
\begin{eqnarray}
\hspace{-0.7in}{\tilde \psi}({\bf k})&=&\rho \pi^{d/2} a^d\exp[-(ka)^2/4] \nonumber\\
&&\times \Bigg[ 1-\Gamma(1+d/2) 
\left(\frac{2}{kD}\right)^{d/2}
\left(\frac{\rho}{\rho_c}\right) J_{d/2}(kD)\Bigg].
\label{PSI}
\end{eqnarray}
Substituting  this expression into (\ref{chi-field}) with $\rho=\rho_c$ and expanding the spectral density in powers of $k^2$ about the
origin yields 
\begin{equation}
{\tilde \psi}({\bf k})= \frac{\pi^{d/2} \rho_c a^d}{2(d+2)} k^2 + {\cal O}(k^4).
\end{equation}
Note that this scalar field is hyperuniform such 
that ${\tilde \psi}({\bf k})$  goes to zero quadratically in $k$ as the wavenumber tends to zero,
independent of the space dimension $d$.

\begin{figure}[bthp]
\begin{center}
\includegraphics[  width=3.in, keepaspectratio,clip=]{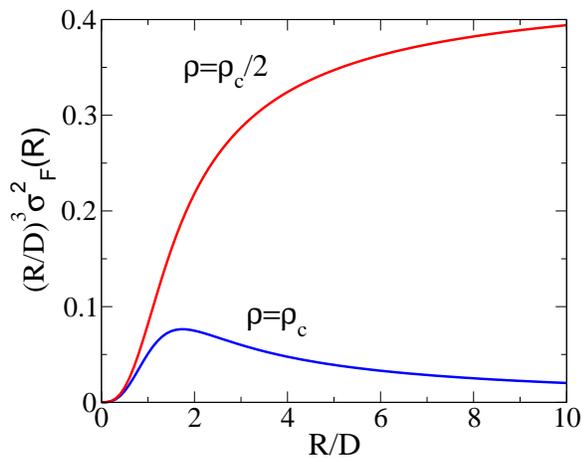}
\caption{ (Color online) Comparison of the field variance $\sigma^2_{F}(R)$ [multiplied by $(R/D)^3$]
versus window sphere radius $R/D$ for the three-dimensional Gaussian field derived from   
the step-function $g_2$-invariant packing for a nonhyperuniform case ($\rho=\rho_c/2$) and 
the hyperuniform case ($\rho=\rho_c$). Here $\rho_c=3/(4\pi)$  and $a=D$, where $D$ is a hard-sphere diameter.}
\label{gauss-2}
\end{center}
\end{figure}

 Figure \ref{gauss} shows
this spectral function ${\tilde \psi}({\bf k})$
in the special case of three dimensions at the hyperuniform terminal density
as well as at a nonhyperuniform case. The scaled corresponding variances,
obtained from relations (\ref{local-scalar}) and (\ref{PSI}), are shown in Fig. \ref{gauss-2}.
 Note that since $\sigma^2_{_F}(R)$ for  the non-hyperuniform case must
decay like $R^{-3}$ for large $R$,  the product $R^3 \sigma^2_{_V}(R)$ asymptotes to a constant value.
By contrast, the product  $R^3 \sigma^2_{_F}(R)$ for $\rho=\rho_c$ decays like $R^{-1}$ for large $R$,
as it should for this three-dimensional hyperuniform random scalar field.

\subsection{Level Cuts of Random Fields}\vspace{-0.1in}

In the random-field approach to modeling the microstructure
of random media, the interface between the phases is defined
by level cuts of random fields \cite{Berk87,Be91,Te91,Cr91,Bl93,Ro95}. 
There is great flexibility in the choice of the
random field $F({\bf x})$ and hence in the class of microstructures that can
be produced. This approach is particularly useful in modeling
{\it bicontinuous} media (two-phase media in which each phase percolates),
such as microemulsions \cite{Berk87}, carbonate rocks \cite{Cr91},
Vycor glass \cite{Cr91}, amorphous alloys, \cite{Ro95} and aerogels \cite{Ro97}.
It is noteworthy that the use of  level cuts of random fields 
to create disordered hyperuniform two-phase or multiphase heterogeneous systems
has heretofore not been carried out, and thus represents a fruitful
area for future research. To derive a hyperuniform two-phase
medium from a thresholded random field $F({\bf r})$, the field must possess the special
correlations required to yield an autocovariance function $\chi_{_V}({\bf r})$ that satisfies the 
 rule (\ref{sum-2}).

\section{Divergence-Free Random Vector Fields and Hyperuniformity}\vspace{-0.1in}
\label{vector}

It is natural to generalize the hyperuniformity concept for scalar fields to  
random vector fields. In order to narrow the enormous possibilities in this 
substantially broader context, we will focus primarily on
divergence-free random vector fields, but the basic ideas
apply to more general vector fields. Excellent physical examples within this class of fields
occur in heterogeneous  media, including 
divergence-free heat, current or mass flux fields, divergence-free electric displacement
fields associated with dielectrics, divergence-free magnetic induction fields, and divergence-free low-Reynolds-number velocity
fields \cite{To02a,Sa03}. Incompressible  turbulent flow fields provide yet
other very well-known set of examples \cite{Ba59,Mo75}.
Here, we derive the relevant equations to quantify
hyperuniform vector fields, present
illustrative calculations, and make contact with turbulent-flow spectra.

Consider a statistically homogeneous divergence-free (solenoidal) random vector field ${\bf u}({\bf x})$ in $\mathbb{R}^d$ 
that is real-valued with zero mean, i.e.,
\begin{equation}
\nabla \cdot {\bf u}({\bf x})=0,
\label{div}
\end{equation}
where
\begin{equation}
\langle {\bf u}({\bf x}) \rangle =0.
\end{equation}
Taking the Fourier transform of (\ref{div}) yields
\begin{equation}
{\bf k} \cdot {\tilde {\bf u}}({\bf k})=0, \qquad \mbox{for all} \; {\bf k},
\end{equation}
where ${\tilde {\bf u}}({\bf k})$ is the Fourier transform of ${\bf u}({\bf x})$.
A key quantity is the autocovariance function $\Psi_{ij}({\bf r})$ ($i,j=1,2,\ldots,d$)
associated with the vector field ${\bf u}({\bf x})$, which is a  second-rank
tensor field defined by
\begin{equation}
\Psi_{ij}({\bf r})= \langle    u_i({\bf x}) u_j({\bf x}+{\bf r}) \rangle,
\label{auto}
\end{equation}
where we have invoked the statistical homogeneity of the field.  
The divergence-free condition (\ref{div})  implies 
\begin{equation}
\frac{\partial \Psi_{ij}({\bf r})}{\partial r_i}=0
\label{1}
\end{equation}
and
\begin{equation}
\frac{\partial \Psi_{ij}({\bf r})}{\partial r_j}=0,
\label{2}
\end{equation}
where the second equation follows from the symmetry property $\Psi_{ij}({\bf r})=\Psi_{ji}(-{\bf r})$
and Einstein indicial summation notation is implied.  Taking the Fourier transforms of (\ref{1}) and (\ref{2}) yield the 
identities
\begin{equation}
k_i {\tilde \Psi}_{ij}({\bf k})= k_j {\tilde \Psi}_{ij}({\bf k})=0, \qquad \mbox{for all} \; {\bf k}.
\label{X}
\end{equation}
where  ${\tilde \Psi}_{ij}({\bf k})$ is the spectral density tensor, i.e., the Fourier transform of 
the autocovariance tensor (\ref{auto}). The real-valued spectral density tensor
is positive semi-definite, i.e., for an arbitrary real vector $\bf a$,
\begin{equation}
a_i {\tilde \Psi}_{ij}({\bf k}) a_j \ge 0, \qquad \mbox{for all} \; {\bf k}.
\end{equation}


From the theory of turbulence of an incompressible fluid \cite{Ba59,Mo75}, it is well known that if an arbitrary
divergence-free vector field ${\bf u}({\bf x})$ is also isotropic, then the spectral
density tensor must take the following general form:
\begin{equation}
 {\tilde \Psi}_{ij}({\bf k})=\left(\delta_{ij} -\frac{k_ik_j}{k^2}\right) {\tilde \psi}(k),
\label{spec-tensor}
\end{equation}
where $\delta_{ij}$ is the Kronecker delta or identity tensor,
and ${\tilde \psi}(k)$ is a nonnegative scalar radial function of the wavenumber $k=|\bf k|$.
A random vector field is isotropic if all of its associated $n$-point correlation
functions are independent of translations, rotations and reflections of the
coordinates. Note that the trace of ${\tilde \Psi}_{ij}({\bf k})$ is trivially related
to ${\tilde \psi(k)}$, i.e.,
\begin{equation}
{\tilde \Psi}_{ii}({\bf k})=(d-1) {\tilde \psi}(k),
\end{equation}
and so we see that
\begin{equation}
{\tilde \Psi}_{ii}({\bf k}={\bf 0})=(d-1) {\tilde \psi}(k=0)=\int_{\mathbb{R}^d} \Psi_{ii}({\bf r}) d{\bf r}
\end{equation}
and
\begin{equation}
\Psi_{ii}({\bf r}={\bf 0})=\frac{(d-1)}{(2\pi)^d} \int_{\mathbb{R}^d} {\tilde \psi}(k) d{\bf k}.
\end{equation}

Now if the radial function ${\tilde \psi}(k)$ is continuous but positive at $k=0$ (not hyperuniform), it immediately follows from 
the form (\ref{spec-tensor}) that the spectral tensor can only be hyperuniform
in {\it certain directions}. For example, the component ${\tilde \Psi}_{11}({\bf k})$ is
zero for $k=k_1$ (all wave vectors along the $k_1$-axis) and the component ${\tilde \Psi}_{12}({\bf k})$ is
zero whenever $k_1=0$ or $k_2=0$. The fact that the value of  ${\tilde \Psi}_{11}({\bf k})$ 
depends on the direction in which the origin is approached means that it is nonanalytic at $\bf k=0$.
On the other hand, if ${\tilde \psi}(k)$ is hyperuniform and continuous at $k=0$,
then each component of ${\tilde \Psi}_{ij}({\bf k})$ will inherit the radial hyperuniformity
of  ${\tilde \psi}(k)$, and hence is independent of the direction in which
the origin is approached. For example, consider the situation in which ${\tilde \psi}(k)$
admits the following small-wavenumber expansion
\begin{equation}
{\tilde \psi}(k) = a_1 |{\bf k}|^\alpha + {o}(|{\bf k}|^\alpha),
\label{exp}
\end{equation}
where $\alpha$ is a positive constant and $o$ signifies higher order terms. Note that whenever $\alpha$ is a noninteger 
or odd integer, ${\tilde \psi}(k)$ is a nonanalytic function at the origin due to a derivative discontinuity. 
(An analytic radial function would admit an expansion in even powers of the wavenumber only.) 
For any $\alpha >0$, substitution of  (\ref{exp}) in $(\ref{spec-tensor})$ reveals that the
spectral tensor is radially hyperuniform near ${\bf k=0}$ such that  it vanishes as $|{\bf k}|^\alpha$.

We conclude that we need an even more general hyperuniformity concept
in the case of a spectral tensor, namely, one in which hyperuniformity
depends on the direction in which the origin is approached in Fourier
space. Let ${\bf k}_Q$ represent a  $d$-dimensional unit vector emanating
from the origin $\bf k=0$.
We say that the field is  hyperuniform  for a particular component $i=I$ and $j=J$ of the spectral
tensor of a vector field (isotropic or not) in the direction ${\bf k}_Q$ if 
\begin{equation}
\lim_{t \rightarrow {0}} {\tilde \Psi}_{IJ}(t{\bf k}_Q)={\bf 0},
\label{HYPER}
\end{equation}
where $t$ is a scalar parameter.
Note that there are many different unit vectors (directions) for a particular spectral tensor that can satisfy
this condition, whether this set is countable, or it is uncountable because
these unit vectors can occur in a continuous range of directions. Moreover, if the condition
applies independent of the direction of the unit vector,  then 
it reduces to the standard spectral definition of hyperuniformity.

\begin{figure}[bthp]
\begin{center}
\includegraphics*[  width=2.8in,clip=keepaspectratio]{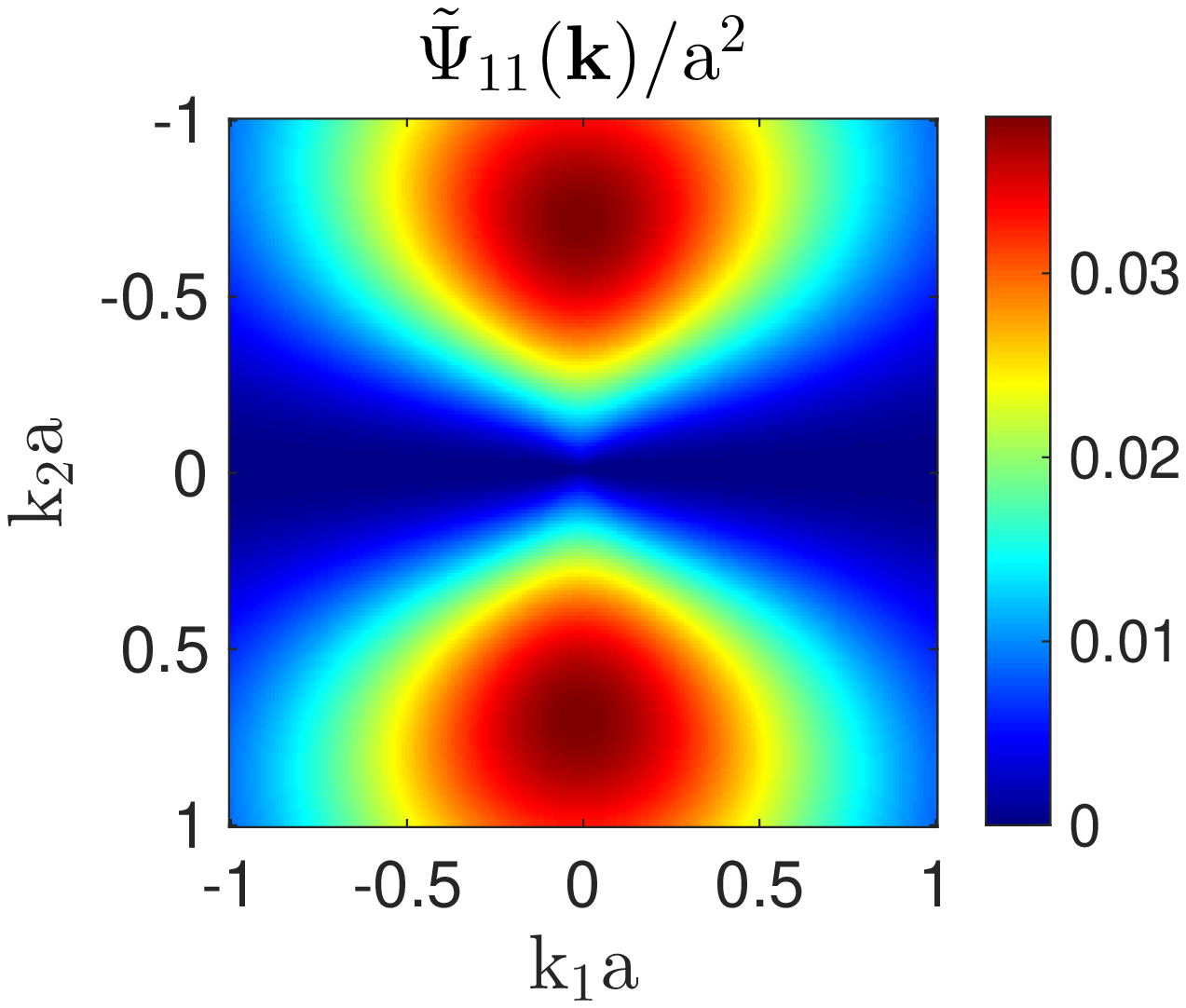}
\includegraphics*[  width=2.8in,clip=keepaspectratio]{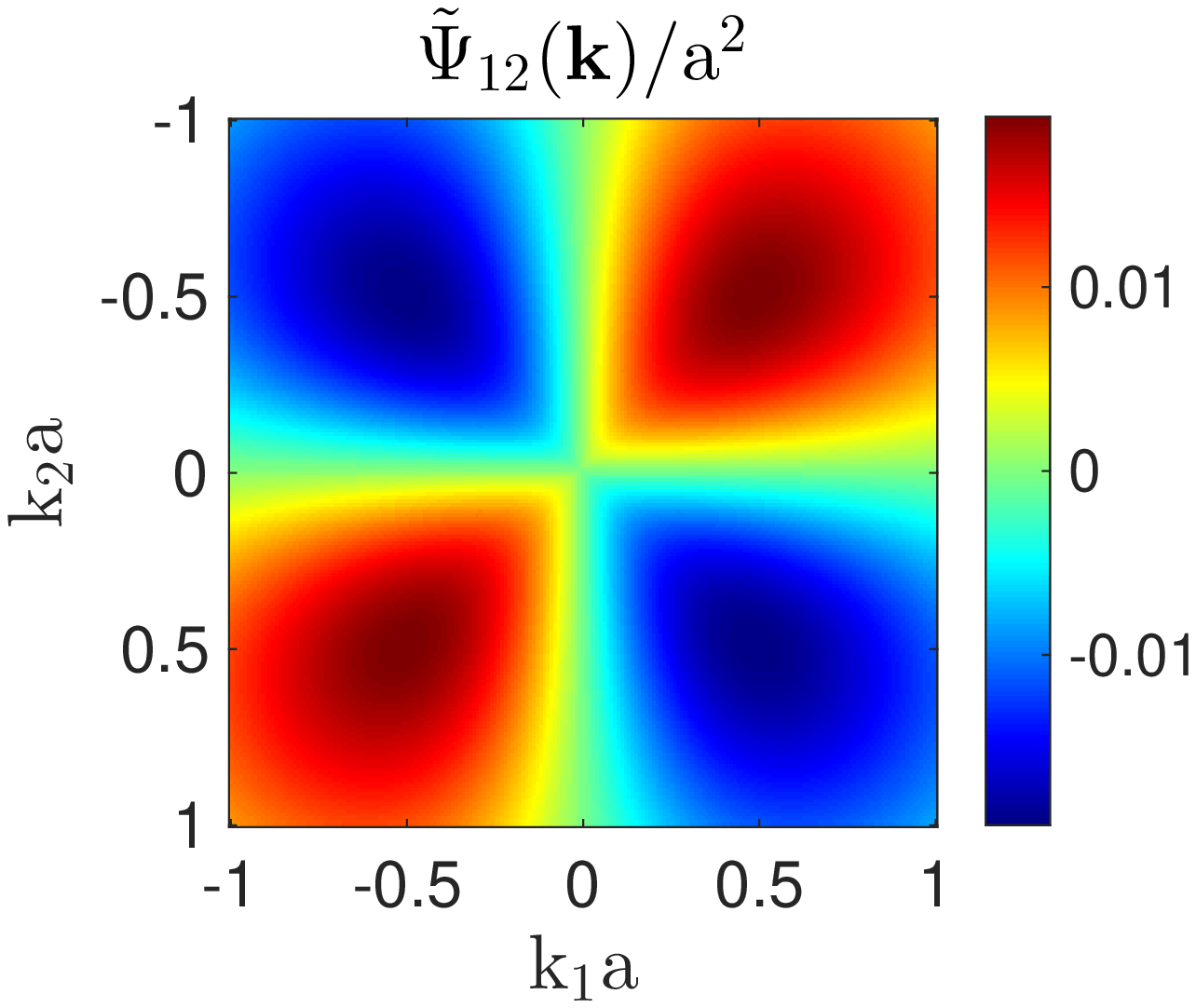}
\includegraphics*[  width=2.8in,clip=keepaspectratio]{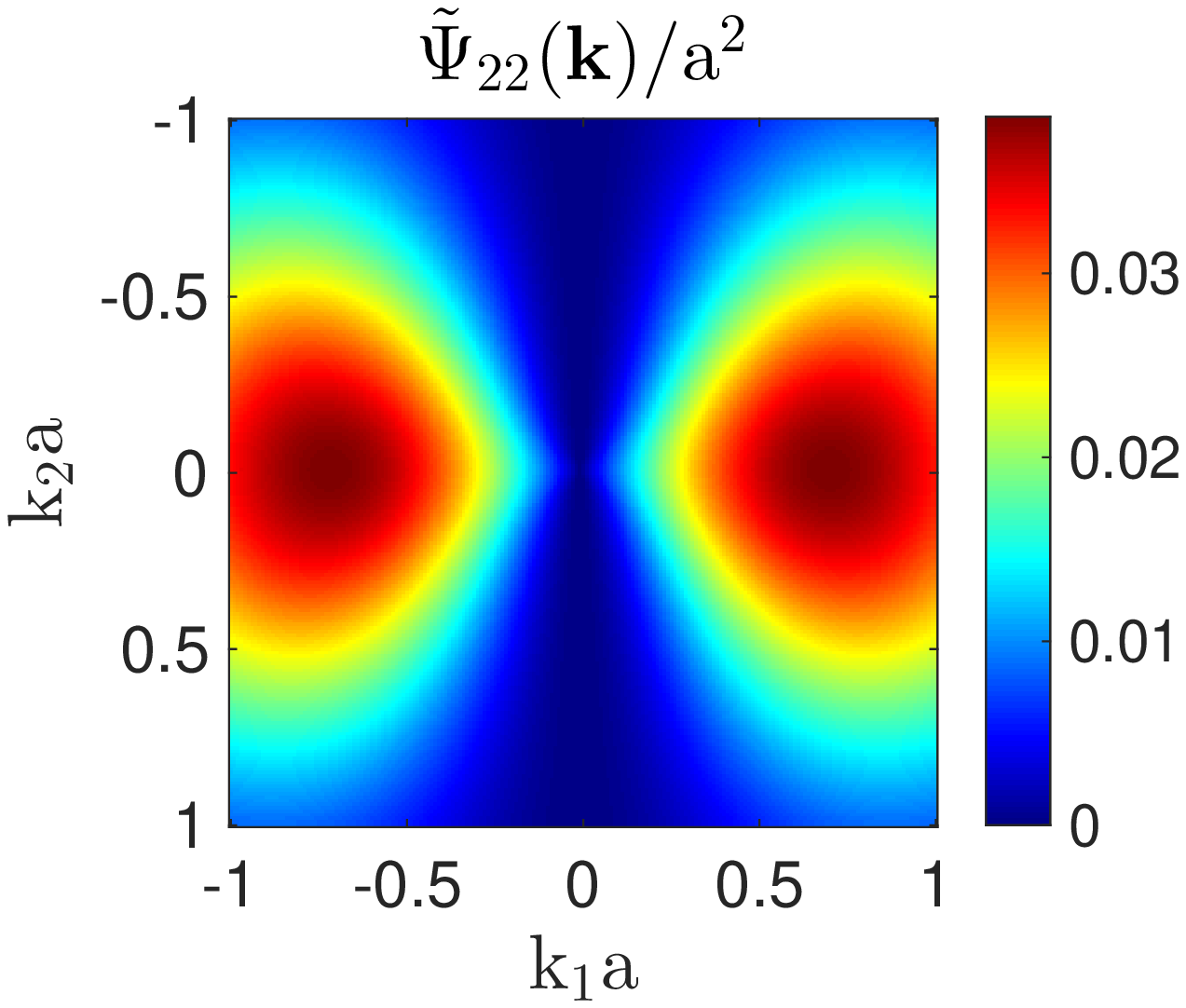}\vspace{0.2in}
\includegraphics*[  width=2.5in,clip=keepaspectratio]{fig7d.eps}
\caption{(Color online) Spectral patterns for the tensor components
of a divergence-free isotropic vector field in $\mathbb{R}^2$ generated
from the radial function (\ref{radial}) with $d=2$, depicted in the bottom panel. Note that unlike the nonnegative 11- and 22-components, 
the 12-component can be both positive and negative, and so its color map
indicating zero intensity (darkest shade) is different from those for the diagonal components.}
\label{tensor}
\end{center}
\end{figure}

To illustrate the hyperuniformity concept in the context of a divergence-free isotropic
vector field, let us consider the
following hyperuniform radial function: 
\begin{equation}
{\tilde \psi}(k)= c(d) (ka) \exp(-(ka)^2),
\label{radial}
\end{equation}
where
\begin{equation}
c(d)= \frac{\Gamma(d/2)a^d}{2^d \pi^{d/2} \Gamma((d+1)/2)}.
\end{equation}
This is a valid (nonnegative) spectral function in any dimension with
an associated autocovariance function $\psi(r)$ such that $\psi(r=0)=1$.
For visual purposes, we examine the two-dimensional outcome when (\ref{radial}) is substituted into
the spectral tensor (\ref{spec-tensor}).  Figure \ref{tensor} shows three components of this
symmetric tensor and the radial function ${\tilde \psi}(k)$. The hyperuniformity
property in a compact region around the origin for all
components is readily visible.

It is instructive to place some well-known results from the theory
of isotropic turbulence in the context of the generalization of  hyperuniformity
to divergence-free random vector fields. For three-dimensional incompressible
turbulent flow with an isotropic velocity field, the radial function ${\tilde \psi}(k)$
appearing in (\ref{spec-tensor})
is simply related to the so-called {\it energy spectrum} of the velocity field, $E(k)$,
via the expression ${\tilde \psi}(k)=E(k)/(4\pi k^2)$.
Thus, we see from the analysis given above that if $E(k)$ goes to zero 
faster than $k^2$ in the limit $k \rightarrow 0$, then each component of the
spectral tensor  ${\tilde \Psi}_{ij}({\bf k})$ will inherit the radial hyperuniformity
of  ${\tilde \psi}(k)$, and hence is independent of the direction in which
the origin is approached. An example of such  energy spectra is one due to 
Batchelor \cite{Ba59}, where $E(k) \sim k^4$ or ${\tilde \psi}(k) \sim k^2$ in the small wavenumber limit.  
Note that the corresponding radial autocovariance function $\psi(r)$ decays
to zero for large $r$ exponentially fast.
On the other hand, if the energy spectrum goes to zero like  $k^2$ or slower in the limit $k \rightarrow 0$,
then the value of the spectral tensor will be hyperuniform only in special
directions. An example within the class of energy spectra is one due to 
Saffman \cite{Sa67}, where $E(k) \sim k^2$ or ${\tilde \psi}(k) \sim \mbox{constant}$ in the small wavenumber limit. 
Here ${\tilde \Psi}_{ij}({\bf k})$ is nonanalytic at $\bf k=0$.
Of course, the significance of energy spectra in turbulence {\it vis a vis}
hyperuniformity was previously not discussed.

\section{Structural Anisotropy and Hyperuniformity}\vspace{-0.1in}
\label{aniso}

\begin{figure}[bthp]
\begin{center}
{\includegraphics[  width=3.3in, keepaspectratio,clip=]{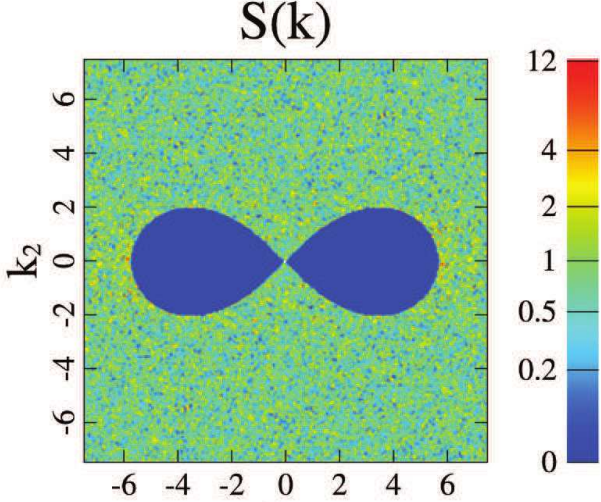}\vspace{-0.4in}
\includegraphics[  width=3.3in, keepaspectratio,clip=]{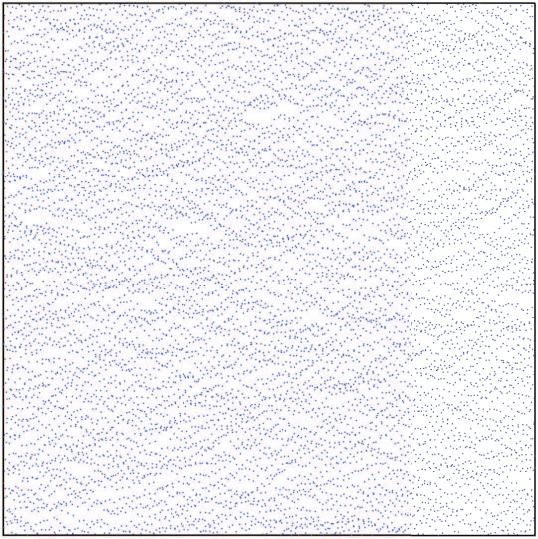}}\vspace{-0.4in}
\caption{(Color online) Top panel: A targeted scattering pattern showing a lemniscate region around 
the origin in which the scattering intensity is exactly zero (darkest shade). This ``stealthy" pattern clearly shows
that hyperuniformity depends on the direction in which the origin $\bf k=0$
is approached. Bottom panel: A statistically anisotropic ground-state configuration of 10,000 particles 
that corresponds
to the unusual scattering pattern shown in the top panel, which is generated using the collective-coordinate
optimization procedure \cite{Uc04b,Zh15a,Zh15b} in a square simulation box under periodic boundary conditions .}
\label{lemniscate}
\end{center}
\end{figure}

\begin{figure}
\begin{center}
\includegraphics[  width=2.5in, keepaspectratio,clip=]{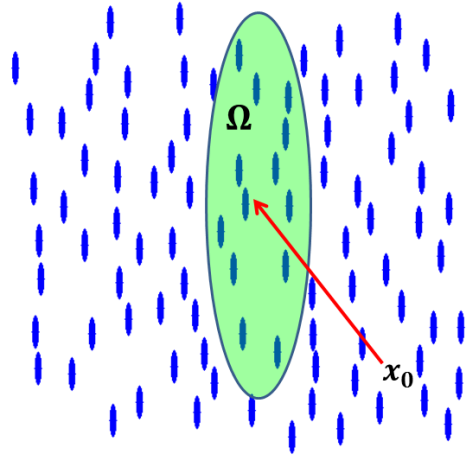}
\caption{(Color online) Schematic illustration of a statistically homogeneous and anisotropic nematic liquid crystal configuration. 
An appropriately shaped window that occupies region $\Omega$ is also shown. Here $\bf x_0$ denotes
both the centroidal position and orientation of the window, the latter of which
is chosen generally from a prescribed probability distribution that depends on the specific
structure of interest. }
\label{nematic}
\end{center}
\end{figure}

Other classes of disordered systems in which ``directional" hyperuniformity is relevant
include many-particle and heterogeneous systems that are statistically
anisotropic, but otherwise statistically homogeneous; see  Figs. \ref{lemniscate} and  \ref{nematic}
for two illustrations. In such cases,
the spectral function conditions (\ref{hyper}), (\ref{hyper-2}) and (\ref{hyper-3}) should be replaced
with the following ones, respectively:
\begin{equation}
\lim_{t \rightarrow 0} S(t{\bf k}_Q) = 0,
\label{Hyper}
\end{equation}
\begin{eqnarray}
\lim_{t \rightarrow 0}\tilde{\chi}_{_V}(t\mathbf{k}_Q) = 0,
\label{Hyper-2}
\end{eqnarray}
\begin{eqnarray}
\lim_{t \rightarrow 0}\tilde{\chi}_{_S}(t\mathbf{k}_Q) = 0,
\label{Hyper-3}
\end{eqnarray}
where the vector ${\bf k}_Q$ is defined in Sec. \ref{vector}.

Are structurally  anisotropic  configurations  associated with 
such exotic spectral functions  realizable?
To vividly demonstrate that the answer to this question is in the affirmative, 
the  collective-coordinate optimization scheme \cite{Uc04b,Ma13,Zh15a,Zh15b} is employed
to produce a many-particle system that is hyperuniform in only certain
directions in Fourier space. This powerful procedure by construction enables 
the structure factor to be constrained to take exact targeted values at a subset of wave vectors.
Whenever the structure factor is constrained to be exactly zero 
for this subset of wave vectors, the resulting configuration exactly corresponds  
to the classical ground state of a long-ranged but bounded pair interaction \cite{Note5}.
For example, one can target stealthy and hyperuniform structure factors that vanish in  a
spherical region around the origin (as in Fig. \ref{pattern}) such that the
associated disordered particle configurations are statistically homogeneous and isotropic ground states \cite{Uc04b,Zh15a,Zh15b}.
Targeted anisotropic structure factors have been attained that correspond to statistically anisotropic ground-state
structures with directional pair interactions \cite{Ma13}, but none of the specific targets computed there were hyperuniform. Here we target a lemniscate region
around the origin $\bf k=0$ in Fourier space in which scattering is completely
suppressed, i.e., this entire region is stealthy, but hyperuniform in only certain
directions; see the top panel of Fig. \ref{lemniscate}. The corresponding
disordered ground states are due to directional long-ranged pair interactions that are stronger
in the horizontal direction than in the vertical  direction, and hence are
characterized by like-linear ``filamentary"  chains of particles that run more or less horizontally.
Such an example is shown in the bottom panel of Fig. \ref{lemniscate}. 

These ground states are characterized by directional-dependent physical properties, including optical, acoustic and elastic behaviors.
Interestingly, although such anisotropic ground-state configurations cannot support shear (for similar reasons
as in their isotropic counterparts \cite{Zh15b}), they are generally elastically anisotropic
because the stress tensor is asymmetric,  as will be detailed in a future study.
In particular, the asymmetry of the stress tensor is associated with internal force couples that resist out-of-plane torques. 
While such behavior is known to occur in liquid crystals and predicted by continuum elastic
theories \cite{De95}, our results are distinguished because the asymmetry of the stress
tensor arises from a microscopic statistical-mechanical model of interacting {\it structureless (point)}
particles. To our knowledge, such a microscopic model has heretofore not been identified.

\begin{figure}[bthp]
\centerline{\includegraphics[  width=2.2in, keepaspectratio,clip=]{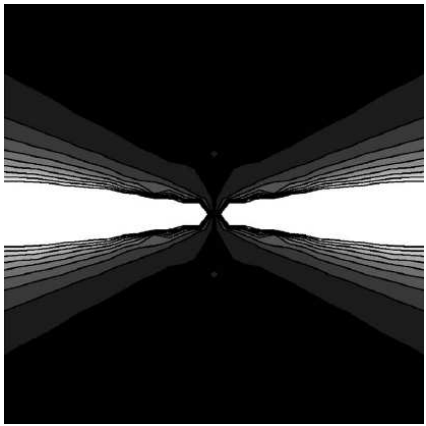}}
\caption{Anisotropic structure factor of a colloidal ferrofluid in the plane in which the particle chains
align, as obtained from  Fig. 6 of Ref. \cite{Hu05}. Dark and light regions indicate low and high intensities,
respectively. Note that depending on the direction in which the origin
is approached, the structure factor can exhibit effective hyperuniformity.}
\label{ferro}
\end{figure}

Many-particle systems that respond to external fields are often characterized
by anisotropic structure factors and hence  provide a class of systems 
where directional hyperuniformity can potentially arise. Huang, Wang and Holm \cite{Hu05} have carried out molecular dynamics
simulations of colloidal ferrofluids subjected to external fields that  capture the salient
structural features observed in corresponding experimental systems as measured
by the structure factor.  Structural anisotropy 
arises in these systems due to the formation of particle chains that tend to align in the direction 
of the applied magnetic field. Figure \ref{ferro} shows an anisotropic structure factor
taken from Ref. \cite{Hu05}. It is apparent that depending on the direction in which the origin
is approached, the structure factor can exhibit effective hyperuniformity.

We can generalize the expressions for the number variance for point configurations and variances for 
a structurally anisotropic two-phase medium
by replacing spherical windows  with an appropriately shaped nonspherical
window occupying region $\Omega$ with an orientation distribution that maximizes sensitivity to direction.
This general formulation was  given in Ref. \cite{To03a} for point configurations,
but no explicit calculations were presented.  Figure \ref{nematic} schematically depicts
a  statistically homogeneous, anisotropic nematic liquid crystal configuration
of particles and an appropriate window shape and orientational distribution to distinguish ``directional" fluctuations associated 
with either the centroidal positions, volume fraction,  or interfacial
area of the particles. It is clear that window sampling in the direction indicated
in Fig. \ref{nematic} will produce fluctuations that are different from those obtained by sampling in the
orthogonal direction. 

Note that the volume-fraction formulas for the autocovariance $\chi_{_V}({\bf r})$
and spectral density ${\tilde \chi}_{_V}({\bf k})$ for sphere packings presented in Sec. \ref{packing-1}  apply as well to the more general class of packings of oriented nonspherical particles by a simple replacement of the
spherical particle  indicator function (\ref{indicator}) with the following one  for a  
nonspherical particle that occupies a region $\omega$:
\begin{equation}
m_v({\bf r};{\bf a}) =\Bigg\{{1, {\bf r}  \in \omega,\atop{0, {\bf r} \notin  \omega,}}
\label{indicator-2}
\end{equation}
where the vector $\bf r$ emanates from the particle centroid and  the vector $\bf a$ represents the set of parameters that defines the shape of the particle.
For example, for a $d$-dimensional ellipsoid, this is given explicitly by
\begin{equation}
m_v({\bf r};{\bf a}) =\Bigg\{{1, \;\;\frac{r_1^2}{a_1^2} +\frac{r_2^2}{a_2^2}+\cdots +\frac{r_d^2}{a_d^2} \le 1,\atop{\hspace{-0.75in}0, \;\; \mbox{otherwise},}}
\label{indicator-3}
\end{equation}
where $r_i$ ($i=1,2,\ldots,d$) is the $i$th the Cartesian component of $\bf r$ and 
$a_1,a_2,\ldots,a_d$ are the semi-axes of the ellipsoid. Of course, the structural
anisotropy for configurations of oriented particles of general shape is reflected in a total correlation
function $h({\bf r})$ or an autocovariance $\chi_{_V}({\bf r})$ that depends not only on the magnitude but direction of $\bf r$.
Observe also that the calculation of $h({\bf r})$ and $\chi_{_V}({\bf r})$ for the special case
of oriented ellipsoids is greatly simplified by exploiting the fact that an ellipsoid
is an affine scale transformation of a sphere  \cite{Le83,La90}.

Similarly, the surface-area formulas for the autocovariance $\chi_{_S}({\bf r})$ and spectral density ${\tilde \chi}_{_S}({\bf k})$
for sphere packings presented in Sec. \ref{packing-2}  still apply
to packings of oriented nonspherical particles when the radial functions $m_s(r;a)$ are replaced with the appropriate
vector-dependent interface indicator function for a particle $m_s({\bf r};{\bf a})$, which is a generalized function that has measure
only on the particle surface. As in the case of anisotropic point configurations, the variances for both volume fraction
and surface area, $\sigma_{_V}^2(R)$ and $\sigma^2_{_S}(R)$, for sphere packings using spherical
windows of radius $R$ can be generalized to allow for anisotropic packings of nonspherical
particles with an appropriately shaped nonspherical window \cite{To03a}.

\section{Conclusions and Discussion}\vspace{-0.1in}
\label{con}

We have generalized the hyperuniformity concept in four different directions:
(1) interfacial area fluctuations in heterogeneous materials; (2) random scalar fields;
(3) divergence-free random vector fields; and (4) statistically anisotropic many-particle
systems and heterogeneous media. These generalizations provide theoreticians and experimentalists new research avenues to understand a very broad 
range of phenomena across a variety of fields through the hyperuniformity ``lens."

The surface-area variance $\sigma_{_S}^2(R)$ and associated
spectral density function ${\tilde \chi}_{_S}({\bf k})$  could play a new and major role in characterizing the
microstructure of two-phase systems, including fluid-saturated porous media, physical
properties that intimately depend on the interface geometry, such
as reaction rates and fluid permeabilities \cite{To02a},
and evolving  microstructures that depend on interfacial
energies (e.g., spinodal decomposition). It should not go unnoticed that the hyperuniformity
concept for two-phase media specified by the volume-fraction and surface-area
variances $\sigma_{_V}^2(R)$ and  $\sigma_{_S}^2(R)$, respectively, are fluctuations
that describe two of the Minkowski functionals  \cite{Sc11}.
In the case of sphere packings, we showed that the surface-area spectral function exhibits stronger and longer-ranged correlations 
compared to the volume-fraction spectral function, indicating that the former
is a more sensitive microstructural descriptor.


Little is known about the hyperuniformity of random scalar fields
and its potential significance. Now that we know what to look for
in such contexts, exploration of this uncharted territory
may prove to be profitable. For example, one could imagine designing random scalar fields
to be hyperuniform (e.g., laser speckle patterns) for photonics
applications \cite{Wi08,Dib16}. 

Our generalization of the hyperuniformity concept to random vector fields
is the most encompassing to date. This setting generally involves a spectral density tensor,
which of course contains random scalar fields as special cases. Even the restricted class of divergence-free
vector fields that we focused on here revealed the need to extend
the ``isotropic" hyperuniformity notion, since the spectral tensor is nonanalytic
at zero wave vector, i.e., it depends on the direction in which the origin in Fourier space
is approached.  Among other results,
we placed well-known energy spectra from the theory
of isotropic turbulence in the context of this generalization of  hyperuniformity. More generally, our work provides a motivation to design 
random vector fields  with targeted directional hyperuniform spectra,
which heretofore has never been considered.

Structurally anisotropic many-particle and heterogeneous systems can also possess
directional hyperuniformity. To illustrate the implications of this generalization,
we presented a disordered directionally hyperuniform many-particle configuration 
that remarkably is the ground state associated with a bounded anisotropic pair potential;
see Fig. \ref{lemniscate}. These filamentary-like ground-state configurations
 will be characterized by directional-dependent physical properties, including optical and elastic behaviors.
Interestingly, such anisotropic ground-state configurations  generally will possess 
internal force couples that resist out-of-plane torques, which will be shown in detail elsewhere. 
Based on our previous investigations using disordered isotropic ground-state configurations
to produce disordered dielectric network solids with large isotropic band gaps \cite{Fl09b,Man13b},
we expect that one can design directional hyperuniform ground-state configurations to 
yield disordered network solids that can be tuned to have photonic and acoustic band gaps with widths
that are relatively uniform for a continuous range of directions and no band gaps
for a different continuous range of directions. Such tunablity could have technological
relevance for manipulating light and sound waves in ways heretofore not thought possible.
Moreover, materials made of dense disordered scatterers 
that are directionally hyperuniform can be designed to be transparent in selected
directions, as a recent study of traditional hyperuniform systems would suggest \cite{Le16}.

Directional structural hyperuniformity raises the interesting possibility that there
may exist disordered many-particle systems in equilibrium that at positive 
temperature $T$ are incompressible in certain directions  and compressible in other directions 
- a highly unusual situation. To understand this proposition, it is 
useful to recall the well-known fluctuation-compressibility theorem for a single-component
many-particle system in equilibrium at number density $\rho$ and temperature $T$:
\begin{equation}
\rho k_B T \kappa_T = \lim_{|{\bf k}| \rightarrow 0} S({\bf k}),
\label{comp}
\end{equation}
where  $\kappa_T$ is the isothermal compressibility.
We see that  in order to have a hyperuniform
system at positive $T$, the isothermal compressibility must be zero; i.e.,
the system must be incompressible \cite{Za11b,To15}. A well-known model 
that exhibits such behavior is the one-component plasma \cite{Ja81}.
However, if the system possesses directional structural hyperuniformity, 
relation (\ref{comp}) no longer applies. Therefore, one must first generalize
this  fluctuation-compressibility theorem to account for directional elastic responses
of the system to different components of stress due to nonanalyticities of
the spectral density at the origin. 
While  (\ref{comp}) has been extended to treat crystals under certain
restrictions \cite{Still66}, to our knowledge, there is
currently no known generalization of (\ref{comp}) that accounts for the anisotropic
elastic response of a disordered equilibrium system to directional stresses
due to nonanalytic spectral densities. Such a generalization of the  fluctuation-compressibility theorem
would enable one to quantify the directions in which the aforementioned hypothesized disordered system is incompressible
or compressible. This represents an intriguing area for future research.
In particular, this possibility challenges experimentalists
to search for such exotic states of matter.

Finally, we note that the hyperuniformity concept has recently been generalized to spin systems,
including a capability to  construct disordered stealthy hyperuniform spin configurations as ground states  \cite{Ch16a}.
The implications and significance of the existence of such disordered
spin ground states warrant further study, including
whether their bulk physical properties and excited states, like their many-particle
system counterparts, are singularly remarkable, and can be experimentally realized.

\vspace{-0.2in}

\acknowledgments{
The author is very grateful to  Duyu Chen, Jaeuk Kim and Zheng Ma               
for their careful reading of the manuscript. He is especially thankful to Duyu Chen and  Ge Zhang
for their assistance in  creating some of the figures.}

\appendix

\section{Multihyperuniformity and Surface-Area Fluctuations in Polydisperse Sphere Packings}\vspace{-0.1in}

Both the autocovariance  function and associated spectral density for packings of hard 
spheres with a continuous or discrete size distribution were previously obtained \cite{Lu91,To02a}.
We collect these results here  to show that when each subpacking associated with each component is
hyperuniform, the entire packing is hyperuniform,
which has been termed {\it multihyperuniformity} in the case of a point configuration \cite{Ji14}.
(The Supplemental Material collects analogous expressions for  $\chi_{_V}({\bf r})$ and ${\tilde \chi}_{_V}({\bf k})$ \cite{Note4}.)

In the case of a continuous distribution in radius ${\cal R}$ is characterized by
a probability density function $f({\cal R})$ that normalizes to unity, 
\begin{equation}
\int_{0}^{\infty} f({\cal R}) d{\cal R}=1.
\label{f-normal}
\end{equation}
Let us denote the size average of a function $G({\cal R})$ by
\begin{equation}
\langle G({\cal R}) \rangle_{\footnotesize{\cal R}} \equiv \int_{0}^{\infty} f({\cal R}) G({\cal R}) d{\cal R}.
\label{size}
\end{equation}
The specific surface and the autocovariance function  are given respectively by
\begin{equation}
s = \rho \langle s_1({\cal R}) \rangle_{\footnotesize{\cal R}}
\end{equation}
and 
\begin{eqnarray}
\chi_{_S}({\bf r})& =&  \rho \langle m_s(r;{\cal R}) \otimes m_s(r;{\cal R}) \rangle_{\footnotesize{\cal R}} \nonumber\\
&+&\rho^2  \Big\langle \Big\langle 
 \,m_s(r;{\cal R}_1) \otimes m_s(r;{\cal R}_2) \otimes h({\bf r};{\cal R}_1,{\cal R}_2) \Big\rangle_{\footnotesize{\cal R}_1} \Big\rangle_{\footnotesize{\cal R}_2}, \nonumber\\
\label{poly}
\end{eqnarray}
where $h({\bf r};{\cal R}_1,{\cal R}_2)$ is the appropriate generalization
of the total correlation function for the centers of two spheres of radii ${\cal R}_1$ and ${\cal R}_2$ separated
by a distance $r$. Note that generally $h$ is not symmetric with respect
to interchange of the components, i.e., $h({\bf r};{\cal R}_1,{\cal R}_2) \neq h({\bf r};{\cal R}_2,{\cal R}_1)$.
Fourier transformation of (\ref{poly}) gives the corresponding
surface-area spectral density
\begin{eqnarray}
\hspace{-0.2in}{\tilde \chi}_{_S}({\bf k}) & =& \rho   \langle {\tilde m}_s^2(k;{\cal R}) S({\bf k};{\cal R})\rangle_{\footnotesize{\cal R}} \nonumber\\
&+&   \rho^2 \Big\langle  \Big\langle {\tilde m}_s(k;{\cal R}_1)  {\tilde m}_s(k;{\cal R}_2)  {\tilde h}({\bf k};{\cal R}_1,{\cal R}_2)  \Big\rangle_{\footnotesize{\cal R}_1}  \Big\rangle_{\footnotesize{\cal R}_2}\nonumber\\
&-&  \rho^2\langle {\tilde m}_s^2(k;{\cal R}) {\tilde h}({\bf k};{\cal R})\rangle_{\footnotesize{\cal R}},
\label{poly-2}
\end{eqnarray}
where 
\begin{equation}
 S({\bf k};{\cal R})=1+\rho {\tilde h}({\bf k};{\cal R}) 
\end{equation}
is the nonnegative structure factor for particles of radius $\cal R$. 
While the first term on the right side of relation (\ref{poly-2}) 
must be nonnegative for all $\bf k$, the remaining two terms together can be negative for some ${\bf k}$. 
The hyperuniformity condition  is obtained by evaluating
(\ref{poly-2}) and setting it equal to zero \cite{Note6}, i.e.,
\begin{eqnarray}
{\tilde \chi}_{_S}({\bf 0}) & =&  0=\rho   \langle s^2({\cal R}) S({\bf 0};{\cal R})\rangle_{\footnotesize{\cal R}} \nonumber\\
&+&   \rho^2\Big\langle  \Big\langle s({\cal R}_1)  s({\cal R}_2)  {\tilde h}({\bf 0};{\cal R}_1,{\cal R}_2)  \Big\rangle_{\footnotesize{\cal R}_1}  \Big\rangle_{\footnotesize{\cal R}_2}\nonumber\\
&-&  \rho^2\langle s^2({\cal R}) {\tilde h}({\bf 0};{\cal R})\rangle_{\footnotesize{\cal R}}.
\label{poly-3}
\end{eqnarray}

One can obtain corresponding results for  spheres
with $M$ different  radii $a_1,a_2,\ldots,a_M$ from the continuous case \cite{To90d,To02a} by letting 
\begin{equation}
f(R) = \sum_{i=1}^{M} \frac{\rho_{i}}{\rho} \delta ({\cal R} -a_{i}), 
\label{f-discrete}
\end{equation}
where $\rho_{i}$ is the number density 
of type-$i$ particles, respectively, and $\rho$ is the {\it total number density}.
Thus, from the relations above and (\ref{f-discrete}): \vspace{-0.3in}

\begin{equation}
s =\sum_{i=1}^M \rho_i s_1(a_i),
\end{equation}
\vspace{-0.4in}

\begin{eqnarray}
\chi_{_S}({\bf r})  &=& \sum_{i=1}^M \rho_i v_2^{int}(r;a_i)  \nonumber\\
&+&  \sum_{i=1}^M  \sum_{j=1}^M \rho_i \rho_j  
 \,m_s(r;a_i) \otimes m_s(r;a_j) \otimes h({\bf r};a_i,a_j) \nonumber\\
\label{chi-r-dis}
\end{eqnarray}
\vspace{-0.3in}

and
\begin{eqnarray}
{\tilde \chi}_{_S}({\bf k}) &=&\sum_{i=1}^M \rho_i {\tilde m}_s^2(k;a_i)  S({\bf k};a_i) \nonumber\\
&+&  \sum_{i\neq j}^M   \rho_i \rho_j  
 \,{\tilde m}_s(k;a_i) {\tilde m}_s(k;a_j) {\tilde h}({\bf k};a_i,a_j).
\label{chi-k-dis}
\end{eqnarray}
It immediately follows that at the origin ${\bf k =0}$ we have
\begin{eqnarray}
{\tilde \chi}_{_S}({\bf 0}) &=&\sum_{i=1}^M \rho_i s^2(a_i)  S({\bf 0};a_i) \nonumber \\
&+&  \sum_{i\neq j}^M   \rho_i \rho_j  
 \,s(a_i) s(a_j) {\tilde h}({\bf 0};a_i,a_j). 
\label{chi-k-hyp}
\end{eqnarray}

When the spatial patterns associated with each component
of a polydisperse packing are themselves hyperuniform [i.e.,
the first term on the right side of (\ref{chi-k-hyp}) is zero], it follows that the second term must be identically zero,
and hence the polydisperse packing is multihyperuniform with respect to surface-area fluctuations.
The proof follows in exactly the same way as for multihyperuniformity 
of polydisperse sphere packings with respect to volume-fraction fluctuations \cite{To16b},
and hence is not presented here explicitly.

Note that any decoration of a crystal in which each component
is arranged in a periodic fashion is multihyperuniform. By contrast, constructing disordered multihyperuniform polydisperse
packings is considerably more challenging. The photoreceptor mosaics in avian retina
are such examples offered by Nature \cite{Ji14}.

Examining the structure factor $S({\bf k})$ of the point
configurations derived from the centers of spheres in a polydisperse packing could lead one to incorrectly
conclude that the packing is not hyperuniform. One way to ascertain hyperuniformity in this case is through a packing's phase spectral
density ${\tilde \chi}_{_V}({\bf k})$ \cite{Za11a,Za11c,Za11d,Be11}. Another way is through surface-area
spectral density ${\tilde \chi}_{_S}({\bf k})$ via the equations given in this appendix.

\end{document}